\documentclass[12pt,halfline,a4paper]{ouparticle}
\usepackage[round]{natbib}   
\usepackage{xurl}   
\usepackage[hidelinks]{hyperref}  
\hypersetup{  
  pdftitle={Do Female Directors Raise ESG Ratings? A Meta-Analysis},
  pdfauthor={Karolina Hozova, Tomas Havranek, Zuzana Irsova},
  pdfsubject={Meta-analysis of board gender diversity and ESG ratings},
  pdfkeywords={board gender diversity, ESG ratings, meta-analysis, publication bias, Bayesian model averaging}
}
\usepackage{ragged2e}   
\usepackage[bottom]{footmisc}   
\usepackage{threeparttable}
\usepackage{graphicx}
\usepackage{booktabs}
\usepackage{multicol}
\usepackage{adjustbox}
\usepackage{tikz}
\usetikzlibrary{shapes.geometric, arrows, positioning}
\usepackage{float}          
\newsavebox{\prismabox}     

\newcommand{\sym}[1]{\ensuremath{^{#1}}}      
\providecommand{\acs}[1]{\MakeUppercase{#1}}  
\providecommand{\say}[1]{``#1''}              

\begin{document}
\title{Do Female Directors Raise ESG Ratings?\break{A Meta-Analysis\thanks{Corresponding author: Karolina Hozova, Institute of Economic Studies, Faculty of Social Sciences, Charles University, Prague; karolina.hozova@fsv.cuni.cz. The full replication package (data, code, and the online appendix) is available at \url{https://meta-analysis.cz/esg}. Hozova acknowledges support from the European Union's Horizon 2020 research and innovation programme under the Marie Sk{\l}odowska-Curie grant agreement No. 870245.}}}

\author{%
\name{Karolina Hozova$^{a}$, Tomas Havranek$^{a,b,c}$, and Zuzana Irsova$^{a,c}$}
\address{\small{$^{a}$Charles University, Prague \\
$^{b}$Centre for Economic Policy Research, London\\
$^{c}$Meta-Research Innovation Center, Stanford}}
}

\abstract{Appointing more women to corporate boards is widely expected to also raise firms' environmental, social, and governance (ESG) performance. We provide the first meta-analysis of this relationship, drawing on 533 estimates from 106 studies that measure ESG performance with Bloomberg or LSEG ratings. The average reported effect of a one-percentage-point increase in board gender diversity is about 0.28 ESG points, but much of it does not survive scrutiny. Correcting for publication bias with a battery of linear and non-linear methods lowers the effect to between roughly 0.08 and 0.17 points. A best-practice estimate that also imposes sound study design puts it near 0.12 for most of the world, markedly higher for the Middle East, and near $-$0.11 for the Southeast Asian markets that dominate the Asian evidence. The differences that remain across studies are systematic, driven mainly by geography and by the choice of estimation method rather than by the ESG-rating provider or the controls a study includes. Board gender diversity may be well worth pursuing on its own merits, but the evidence that it reliably raises ESG scores is weaker than the published record suggests.}

\date{\today}

\maketitle

\newpage

\section{Introduction}
\label{chap:one}
Companies face pressure on two fronts at once. Investors, regulators, and the public want firms to put more women in the boardroom, and they want firms to do better on environmental, social, and governance (ESG) measures. For a board deciding where to spend its effort, a tempting shortcut suggests itself: if women directors genuinely improve a firm's ESG performance, appointing more of them advances both goals together.

The idea is not far-fetched. A large body of work in psychology and management argues that women bring distinct priorities to the boardroom. Theories of gender socialization hold that women are, on average, more attuned to the welfare of others and less willing to tolerate unethical behavior \citep{Gilligan1977, Boulouta2013, Williams2003}. Carried into corporate leadership, these tendencies are thought to make female directors more responsive to the concerns of employees, communities, and the environment \citep{Bear2010, Harjoto2015, Atif2021}.

The empirical record is far less tidy. \citet{Manita2018}, among the most cited studies on the topic, examined US firms between 2010 and 2015 and found an effect indistinguishable from zero. Three years later, \citet{Shakil2021} revisited the question on US banks with more recent data and reported a clear positive effect of roughly half a point. The two papers were soon cited side by side, and the later paper appeared to overturn the earlier null. Similar disagreements run across regions and sectors, from Latin America \citep{Husted2019} and the Middle East \citep{Issa2022} to energy \citep{Shahbaz2020}, healthcare \citep{Uyar2021}, and extractive industries \citep{Wang2022}. Across the 533 estimates we collect, the raw reported effect of a one-percentage-point increase in board gender diversity ranges from $-2.8$ to $5.9$ ESG points before winsorization.

\begin{figure}[htbp]
\centering
\caption{Heterogeneity in the literature} 
\includegraphics[width=0.8\textwidth]{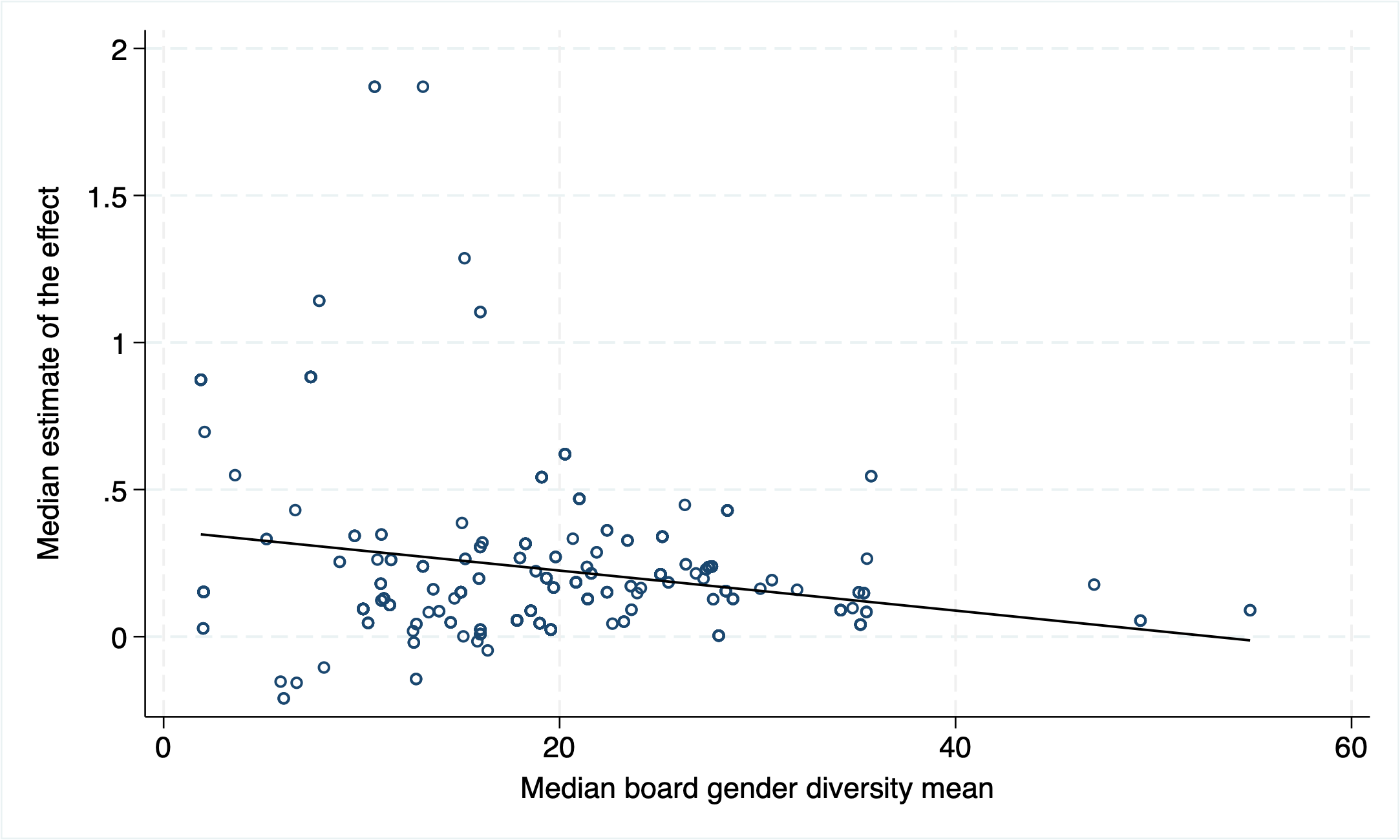}
\label{fig:heterogeneity}
\begin{threeparttable}
    \begin{tablenotes}
        \footnotesize
        \item \textit{Notes:} Each circle represents one primary study. The vertical axis shows the study's median estimate of the effect of a one-percentage-point increase in board gender diversity on ESG scores. The horizontal axis shows the median sample mean of board gender diversity in that study. The fitted line is from an OLS regression.
    \end{tablenotes}
\end{threeparttable}
\end{figure}

Part of this disagreement tracks the level of female board representation itself. Figure~\ref{fig:heterogeneity} plots each study's median effect against the average share of women on its sample's boards. The downward slope is stark: studies from contexts where women hold few board seats (often firms in the Middle East or other emerging markets, where the sample mean rarely exceeds ten percent) report the largest effects, sometimes above one ESG point per percentage point. Studies from high-representation settings such as Northern Europe or the United States cluster near zero. A literature that seems to measure one thing in fact measures very different margins in very different institutional settings. How readily women reach board seats depends on regulation, investor pressure, and cultural norms as much as on the sector, so a single pooled estimate can mask sharply different local relationships \citep{Carrasco2015}.

What explains this spread, and what is the effect beneath it? A literature with an appealing story, mixed findings, and wide variation is fertile ground for publication bias: the tendency for results of the expected sign and conventional significance to be reported, cited, and published more readily than null results \citep{Stanley2005}. The pattern is documented across economics, from minimum-wage research \citep{Card1995, Doucouliagos2009} and fiscal policy \citep{Heinemann2018} to the elasticity of factor substitution \citep{Gechert2022}, the growth effects of remittances \citep{Cazachevici2020}, and behavioral economics \citep{Havranek2017}. The literature on board gender diversity looks especially exposed. A null result on such an appealing hypothesis makes for a far less compelling paper than a positive one, so positive and significant estimates likely reach print more easily. The average effect circulating in the literature may then overstate whatever genuine relationship exists.

We provide the first assessment of how much of this evidence survives such scrutiny. We hand-collect 533 estimates from 106 studies that measure ESG performance with Bloomberg or LSEG ratings, restricting the sample to this common 0--100 scale so that effect sizes are comparable across studies. These ratings are third-party assessments and need not track a firm's underlying conduct, so our results speak to rated ESG performance rather than to behavior we observe directly. To separate any genuine effect from selective reporting, we apply a battery of linear and non-linear corrections: the FAT-PET regression of \citet{Stanley2005, Stanley2008}, the weighted average of adequately powered estimates of \citet{Stanley2017}, the selection model of \citet{Andrews2019}, the stem-based method of \citet{Furukawa2019}, the endogenous kink model of \citet{Bom2019}, and the p-uniform* estimator of \citet{Aert2026}. We then ask what drives the disagreement, using Bayesian model averaging over 24 variables (the standard error and 23 study characteristics) to avoid betting the analysis on a single specification.

The raw literature implies that a one-percentage-point increase in board gender diversity raises ESG scores by about 0.28 points. Much of that reflects selective reporting: correcting for publication bias alone brings the typical effect down to between 0.08 and 0.17 points, depending on the method. A best-practice estimate that additionally imposes sound study design (a panel setup and no publication bias) puts the effect near 0.12 for most of the world and markedly higher for Middle Eastern firms; for the Southeast Asian markets that dominate our Asian evidence, the effect reverses its sign to about $-$0.11. The heterogeneity across studies is systematic rather than random, driven mainly by geography and by a few study characteristics, especially panel-data use and estimation method. The headline magnitude owes more to how the literature was built than to how firms behave.

Related meta-analyses examine women on boards and firm financial performance \citep{PostByron2015}, women directors and corporate social performance \citep{ByronPost2016, wu2022corporate}, and gender diversity and corporate disclosure \citep{aljanadi2025gender}. None of them studies third-party ESG ratings on a common scale; \citet{wu2022corporate}, for instance, synthesize 44 papers on board gender diversity and corporate social responsibility rather than on rated ESG performance. To the best of our knowledge, this is the first meta-analysis of the association between board gender diversity and third-party ESG ratings measured on a comparable Bloomberg or LSEG scale. It is also the first to test this ESG-ratings literature for publication bias and to identify, rather than assume, what drives the disagreement among its estimates.

The remainder of the paper proceeds as follows. Section~\ref{chap:two} describes the dataset of board gender diversity effects. Section~\ref{chap:three} explores publication bias. Section~\ref{chap:four} examines the drivers of heterogeneity. Section~\ref{chap:five} concludes. Appendix~A details how studies were selected for inclusion, Appendix~B and Appendix~C report robustness checks, and Appendix~D documents our use of artificial intelligence.

The data and code behind every table and figure are available in an online appendix at \url{https://meta-analysis.cz/esg}. The package includes the full dataset of 533 estimates with a codebook, the Stata and R scripts, and documentation of the order in which they run, so that our coding decisions and each step of the analysis can be inspected and reproduced. We follow the reporting guidelines for meta-analyses in economics of \citet{Stanley2013guidelines}, \citet{Havranek2020guidelines}, and \citet{Irsova2023}, together with their update for the use of artificial intelligence \citep{Cook2026guidelines} and the guiding principles that accompany it \citep{Cook2026ai}. Appendix~D reports, stage by stage, where artificial intelligence was and was not used.

\section{Data}
\label{chap:two}

We collect 533 estimates of the effect of board gender diversity on ESG scores from 106 research papers. We identify the primary studies using Google Scholar because it searches the full text of articles, not just titles, abstracts, and keywords. To keep the search process replicable, we rely on a single search query. In addition, we go through the reference lists of the studies already included and add relevant papers that the search itself misses (backward snowballing). We do not perform forward snowballing (tracking citations to the identified studies); instead, shortly before finalizing the dataset we ran a second Google Scholar search to capture studies published since the first search. Complete details of the identification strategy and the specific search query are presented in Figure~\ref{fig:prisma} in Appendix~A.

For comparability, we consider only studies that examine the impact of board gender diversity, measured as the ratio of female directors on the board, on ESG scores rated by Bloomberg or London Stock Exchange Group (LSEG), both reported on the same 0 to 100 scale. ESG scores can differ across providers \citep{Dorfleitner2015, Berg2022}; we test below whether the results differ between Bloomberg and LSEG ratings. To keep the effect size comparable across studies, we exclude estimates that omit one of the three ESG pillars, log-transform the ESG score, or discount it by a controversy score \citep{Shakil2021}, keeping a study's comparable ESG-score estimates when it also reports them. Similarly, we exclude studies that use statistical measures such as the Blau or Shannon index \citep{Abdullah2024, Gangi2021} and studies that focus solely on female CEO or chair leadership \citep{Aabo2023}, the proportion of women in top management teams \citep{Fu2023}, dummies for female board representation \citep{Chebbi2022}, or only the number of women on the board \citep{Kravchenko2023}. On the other hand, we do not exclude studies based on their publication form \citep{Stanley2001}, provided they report a measure of uncertainty such as standard errors, confidence intervals, or p-values. Our final sample therefore includes not only standard peer-reviewed journal articles but also working papers and master's theses. Table \ref{tab:primary} presents the final list of primary studies used in our meta-analysis.

\begin{table}[!t]
\centering

\begin{minipage}{0.95\linewidth}
\centering
\caption{Primary studies used}
\label{tab:primary}
\vspace{0.2cm}
\resizebox{\linewidth}{!}{%
\begin{tabular}{lll}
 \toprule
Adamu et al. (2024) & de Klerk \& Singh (2023) & Nadeem et al. (2017) \\
Adeneye et al. (2024) & De Masi et al. (2021) & Nandi et al. (2023) \\
Adiasih \& Lianawati (2018) & Dicuonzo et al. (2024) & Nekhili et al. (2021) \\
Agnese et al. (2024a) & Disli et al. (2022) & Nery \& Morales (2022) \\
Agnese et al. (2024b) & Donkor et al. (2023) & Nicolo et al. (2021) \\
Agustina \& Barokah (2024) & Ellili (2023) & Nicolo et al. (2023) \\
Ahmadi \& Amara (2024) & Fahad \& Rahman (2020) & Nicolo et al. (2024) \\
Al Kurdi et al. (2023) & Gaio \& Goncalves (2022) & Ozturk (2023) \\
Al-Shaer et al. (2024) & Gerged et al. (2023) & Paolone et al. (2024a) \\
Ali \& Firmansyah (2023) & Giannarakis (2013) & Paolone et al. (2024b) \\
Aliani et al. (2024) & Govindan et al. (2021) & Pinheiro et al. (2023) \\
Aliti \& Wen (2023) & Grubler (2024) & Pinheiro et al. (2024) \\
Alkayed et al. (2024) & Gungor \& Seker (2022) & Pirskanen (2023) \\
Alkhawaja et al. (2023) & Halid et al. (2022) & Qureshi et al. (2020) \\
Almaqtari et al. (2023) & Heubeck (2024) & Qureshi et al. (2023) \\
Almaqtari et al. (2024) & Husted \& de Sousa-Filho (2019) & Rella \& L'Abate (2022) \\
Amara \& Ahmadi (2024) & Issa et al. (2022) & Sari \& Fitriani (2023) \\
Amorelli \& Garcia-Sanchez (2023) & Jizi et al. (2022) & Setiani \& Novitasari (2024) \\
Andreassen \& Bukhari (2024) & Kamaludin et al. (2022) & Shahbaz et al. (2020) \\
Arayssi et al. (2016) & Kamran et al. (2023) & Shakil et al. (2021) \\
Arayssi et al. (2020) & Kampoowale et al. (2024) & Sofiati \& Mita (2024) \\
Arayssi et al. (2024) & Khatri (2023) & Temiz \& Acar (2023) \\
Arduino et al. (2024) & Khemakhem et al. (2023) & Toerien et al. (2023) \\
Ben Fatma \& Chouaibi (2021) & Kouki (2023) & Trireksani et al. (2024) \\
Benaguid et al. (2023) & Lavin \& Montecinos-Pearce (2021) & Uyar et al. (2020) \\
Bhatia \& Marwaha (2022) & Lozano \& Martinez-Ferrero (2022) & Uyar et al. (2021) \\
Bigelli et al. (2023) & Makeeva et al. (2022) & Van Hoang et al. (2023) \\
Birindelli et al. (2018) & Manita et al. (2018) & van Zundert (2024) \\
Boukattaya \& Omri (2021) & Marrone et al. (2024) & Velte (2016) \\
Bruna et al. (2021) & Martinez et al. (2020) & Wang et al. (2022) \\
Buallay et al. (2022) & Martinez et al. (2022) & Waterstraat et al. (2021) \\
Chebbi et al. (2020) & Meen (2023) & Wu et al. (2024) \\
Cucari et al. (2018) & Mehmood et al. (2023) & Yadav \& Prashar (2022) \\
Dakhli (2021) & Miranda et al. (2023) & Yarram \& Adapa (2021) \\
Dang et al. (2021) & Monteiro et al. (2024) &  \\
Dang et al. (2023a) & Moussa \& Elmarzouky (2023) &  \\
\bottomrule
\end{tabular}%
} 

\footnotesize {\justifying \textit{Notes:} The table lists all primary studies identified through the search strategy described in Section~\ref{chap:two} (Google Scholar and snowballing). Full bibliographic references for all 106 studies are provided in the replication package and the online appendix (\url{https://meta-analysis.cz/esg}).  
The last study was added on December 16, 2024.\par}

\end{minipage}
\end{table}

Alongside the individual effect estimates and their uncertainty measures from the primary studies in our list, we further hand-collect other variables to examine the systematic heterogeneity among the reported coefficients: estimation and publication characteristics, the design of the analysis, control variables, data features, and spatial variation.

During data collection, we made a few adjustments to ensure that our dataset contains comparable effect estimates. First, some studies explore a non-linear relationship between board gender diversity and ESG scores by including a quadratic term \citep{Birindelli2018}. To handle the presence of two related estimates, we follow the methodology of \citet{Zigraiova2016} and linearize the effect. Second, four studies report standardized effects 
\citep{Dakhli2021, Dang2023b, Kamran2023, Khemakhem2023}. We recompute the standardized estimates from these studies to raw effects using the standard deviation ratio. Third, some studies employ interaction terms between board gender diversity and other variables, such as common law tradition \citep{Alkhawaja2023}, the critical mass of women on the board \citep{Birindelli2018} or ESG controversies \citep{Shakil2021}. In line with the approach by \citet{Cazachevici2020}, we compute the average marginal effects by applying the delta method to derive the corresponding standard errors. Whenever any of the mentioned transformations is applied, we document it in our dataset, so we can exclude such estimates from our robustness checks. For some studies, the lack of summary statistics or uncertainty measures prevents us from applying the delta method or effect standardization \citep{Dang2023a, Nuhu2024}. Also, in cases where the mean value for the variable included in the interaction term is too high and produces extremely large average marginal effects, we exclude the affected estimates from our dataset \citep{Giannarakis2013}. 

A few studies reported p-values or standard errors of zero, which required imputation. For reproducibility, we replace zero standard errors with 0.0004 \citep{Bruna2021, Gerged2023} and zero p-values \citep{Bhatia2022, Martinez2022, Nadeem2017, Setiani2024} or significance levels of 1\% \citep{Miranda2023} with 0.0001. For studies reporting only that an estimate is significant at the 5\% level \citep{Fahad2020, Kamaludin2022}, we set the t-statistic to 1.96, the boundary value for two-tailed significance at the 5\% level; this is the conservative convention, implying the largest standard error consistent with the reported significance. Estimates with imputed standard errors or p-values are also marked in the dataset and excluded from the robustness check.

\begin{figure}[!t]
    \centering
    \caption{Distribution of gender diversity effects}
    \includegraphics[width=\textwidth]{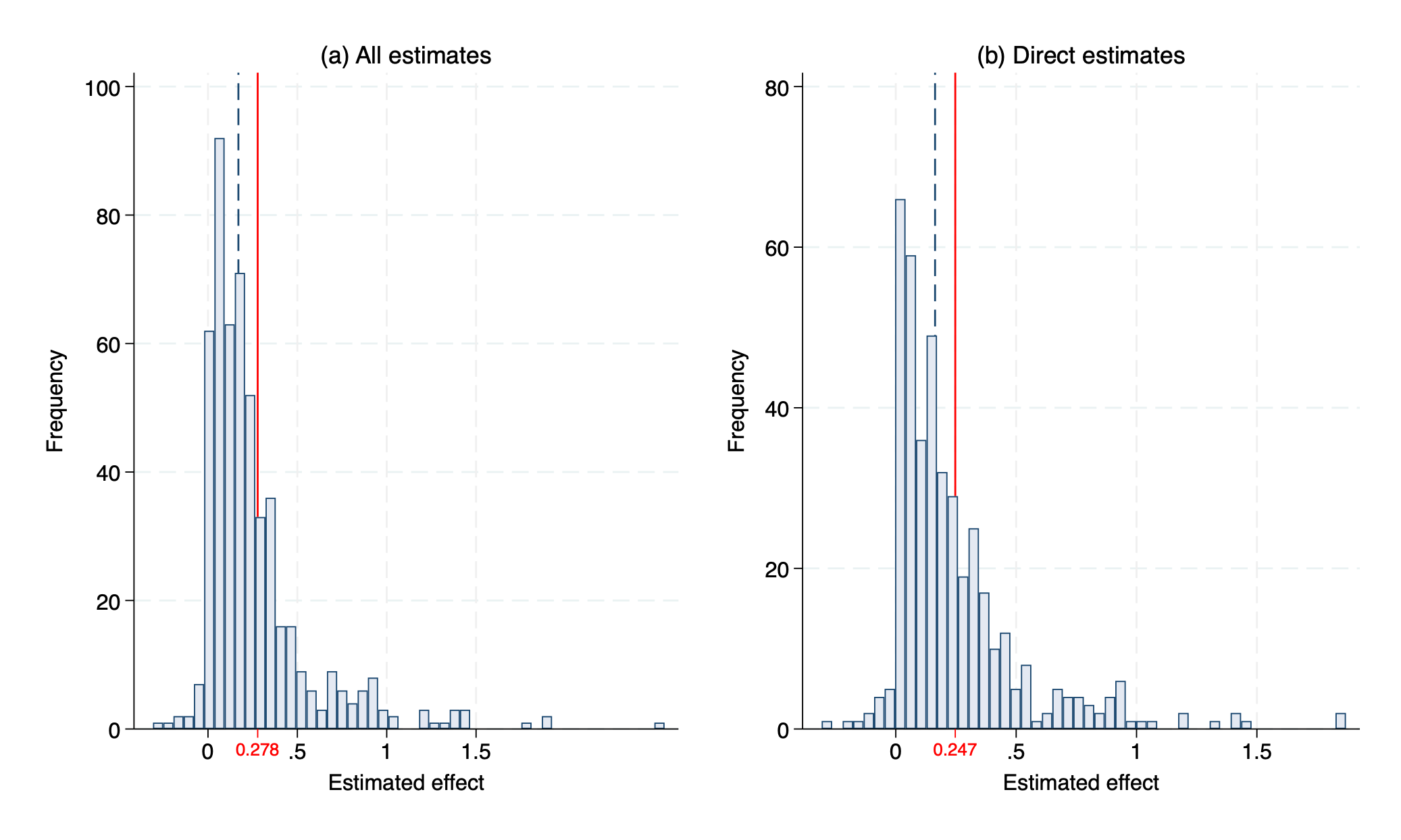}
    \begin{threeparttable}
        \begin{tablenotes}
            \footnotesize
            \item \textit{Notes:} The figure presents a distribution of the estimated effects of board gender diversity on ESG scores as reported across individual studies. Panel~(a) shows all estimates; panel~(b) restricts to direct estimates (those reported without transformation or imputation). For legibility both panels display the range $-0.32$ to $2.73$, which omits eight estimates from panel~(a) and three from panel~(b); the mean lines, the reported summary statistics and every analysis use all estimates. In each panel the solid vertical line marks the mean reported effect and the dashed vertical line the median; across all estimates the mean corresponds to a 0.278-point increase in ESG rating following a one-percentage-point increase in female board representation.
        \end{tablenotes}
    \end{threeparttable}
    \label{fig:estimates_hist}
\end{figure}

\begin{table}[!b]
\centering
\caption{Board gender diversity effects in different contexts}
\label{tab:descr_stat}
\vspace{0.2cm}
\resizebox{\textwidth}{!}{%
\begin{tabular}{l@{\hspace{3em}}c@{\hspace{3em}}cccccc}
\toprule
 & & \multicolumn{3}{c}{Weighted} & \multicolumn{3}{c}{Unweighted} \\
\cmidrule(lr){3-5}\cmidrule(lr){6-8}
 & No.\ of estimates & Mean & \multicolumn{2}{c}{95\% conf.\ int.} & Mean & \multicolumn{2}{c}{95\% conf.\ int.} \\
\midrule

All estimates           & 533 & 0.278 & 0.245 & 0.311 & 0.269 & 0.241 & 0.298 \\
Direct estimates        & 426 & 0.226 & 0.197 & 0.254 & 0.246 & 0.218 & 0.274 \\
Non-direct estimates    & 107 & 0.501 & 0.388 & 0.613 & 0.364 & 0.275 & 0.452 \\

\midrule
\emph{Data type} \\
\hspace*{0.5cm}Data: panel               & 512 & 0.258 & 0.227 & 0.289 & 0.263 & 0.235 & 0.290 \\
\hspace*{0.5cm}Data: cross-sectional    & 21  & 0.544 & 0.262 & 0.825 & 0.434 & 0.144 & 0.723 \\
\hspace*{0.5cm}ESG data: LSEG                    & 329 & 0.297 & 0.258 & 0.336 & 0.270 & 0.236 & 0.303 \\
\hspace*{0.5cm}ESG data: Bloomberg               & 204 & 0.245 & 0.186 & 0.304 & 0.269 & 0.216 & 0.322 \\

\midrule
\emph{Publication status} \\
\hspace*{0.5cm}Published        & 506 & 0.290 & 0.254 & 0.325 & 0.271 & 0.241 & 0.301 \\
\hspace*{0.5cm}Unpublished             & 27  & 0.171 & 0.098 & 0.245 & 0.239 & 0.151 & 0.327 \\

\midrule
\emph{Design of the analysis} \\
\hspace*{0.5cm}Endogeneity control: poor    & 435 & 0.283 & 0.246 & 0.320 & 0.275 & 0.244 & 0.306 \\
\hspace*{0.5cm}Endogeneity control: proper & 98  & 0.259 & 0.182 & 0.335 & 0.245 & 0.172 & 0.318 \\

\midrule
\emph{Control variables} \\
\hspace*{0.5cm}Firm size control: yes           & 474 & 0.299 & 0.263 & 0.335 & 0.284 & 0.253 & 0.316 \\
\hspace*{0.5cm}Firm size control: no            & 59  & 0.123 & 0.067 & 0.179 & 0.149 & 0.099 & 0.199 \\
\hspace*{0.5cm}Board independence control: yes  & 357 & 0.277 & 0.238 & 0.316 & 0.267 & 0.233 & 0.302 \\
\hspace*{0.5cm}Board independence control: no   & 176 & 0.280 & 0.217 & 0.342 & 0.273 & 0.221 & 0.326 \\
\hspace*{0.5cm}CSR committee control: yes       & 206 & 0.234 & 0.195 & 0.273 & 0.259 & 0.216 & 0.301 \\
\hspace*{0.5cm}CSR committee control: no        & 327 & 0.302 & 0.255 & 0.349 & 0.276 & 0.238 & 0.315 \\

\midrule
\emph{Spatial variation} \\
\hspace*{0.5cm}Region: global                  & 177 & 0.290 & 0.226 & 0.354 & 0.219 & 0.167 & 0.270 \\
\hspace*{0.5cm}Region: Europe                  & 182 & 0.202 & 0.181 & 0.224 & 0.232 & 0.208 & 0.257 \\
\hspace*{0.5cm}Region: USA                     & 42  & 0.316 & 0.198 & 0.433 & 0.295 & 0.197 & 0.394 \\
\hspace*{0.5cm}Region: Asia                    & 43  & 0.174 & 0.056 & 0.293 & 0.177 & 0.071 & 0.283 \\
\hspace*{0.5cm}Region: Middle East             & 30  & 0.577 & 0.394 & 0.760 & 0.657 & 0.459 & 0.855 \\
\hspace*{0.5cm}Region: other         & 59  & 0.462 & 0.295 & 0.629 & 0.388 & 0.277 & 0.499 \\
\hspace*{0.5cm}Market: developed             & 274 & 0.274 & 0.234 & 0.314 & 0.257 & 0.229 & 0.286 \\
\hspace*{0.5cm}Market: emerging                & 98  & 0.278 & 0.194 & 0.362 & 0.411 & 0.318 & 0.504 \\
\hspace*{0.5cm}Market: mixed                   & 161 & 0.290 & 0.216 & 0.364 & 0.204 & 0.147 & 0.261 \\
\hspace*{0.5cm}Sector: financial               & 37  & 0.294 & 0.206 & 0.381 & 0.385 & 0.264 & 0.506 \\
\hspace*{0.5cm}Sector: non-financial           & 152 & 0.197 & 0.173 & 0.221 & 0.216 & 0.186 & 0.246 \\
\hspace*{0.5cm}Sector: mixed                   & 344 & 0.313 & 0.264 & 0.363 & 0.281 & 0.240 & 0.321 \\

\midrule
\emph{Estimation techniques} \\
\hspace*{0.5cm}Method: linear       & 186 & 0.263 & 0.215 & 0.311 & 0.264 & 0.220 & 0.307 \\
\hspace*{0.5cm}Method: panel (FE or RE)            & 215 & 0.226 & 0.185 & 0.267 & 0.238 & 0.196 & 0.280 \\
\hspace*{0.5cm}Method: IV   & 33  & 0.656 & 0.437 & 0.875 & 0.423 & 0.253 & 0.592 \\
\hspace*{0.5cm}Method: GMM                     & 42  & 0.334 & 0.167 & 0.502 & 0.282 & 0.141 & 0.424 \\
\hspace*{0.5cm}Method: other         & 57  & 0.224 & 0.139 & 0.308 & 0.309 & 0.219 & 0.399 \\
\bottomrule
\end{tabular}%
}
\begin{minipage}{\textwidth}
\vspace{0.15cm}
\footnotesize
\textit{Notes:} The table displays subgroup summary statistics of the estimated effects of board gender diversity on ESG scores. Weighted means give each study equal weight; unweighted means give each estimate equal weight.
\end{minipage}
\end{table}

Finally, some outliers in the estimated effect sizes and their standard errors survive the cleaning. To keep these observations informative without letting them skew the results, we winsorize the estimates and their standard errors at the 1\% level, replacing the five most extreme observations in each tail of each variable, so that the largest retained effect is about 1.9 ESG points. Appendix~B reports all publication-bias tests on the raw, unwinsorized data. The non-linear and precision-weighted corrections remain small, positive, and statistically significant in this check. But on the raw data the linear FAT-PET slope loses significance in the unweighted and study-weighted specifications, whose corrected means rise to about 0.25--0.31; we therefore rest the corrected-effect claim on the estimators that survive in both samples.

The final dataset includes 533 estimates derived from 106 primary studies. The studies are dated between 2013 and 2024; only 16 of the 106 appeared before 2021, reflecting the topic's recent rise. Figure \ref{fig:estimates_hist} shows the distribution of estimated effects. The left-hand panel shows all estimates, while the right-hand panel restricts to direct estimates (those reported without transformation or imputation). Both histograms show a heavy-tailed distribution with a peak around zero and positive skewness. 

The box plot in Figure \ref{fig:estimates_hbox} in Appendix A illustrates the heterogeneity of the estimates in different regions. Consistent with the histograms presented in Figure \ref{fig:estimates_hist}, most of the estimates are positive. Some regions, notably Latin America and the Middle East, exhibit greater variation, with wider interquartile ranges and longer whiskers. 

Table \ref{tab:descr_stat} puts numbers on these patterns, reporting both weighted and unweighted means across categories. A weighted mean gives each study equal total weight (the inverse of its number of estimates); an unweighted mean gives each estimate equal weight. We prefer weighted means because our dataset includes primary studies that report disproportionately many estimates \citep{Alkhawaja2023}, whereas other primary studies report only a single estimate \citep{Waterstraat2021, Yarram2021}. Although the overall sample means do not differ dramatically, some subsample means vary much more. The overall estimated effect of female board participation on firms' ESG is around 0.278 (weighted) and 0.269 (unweighted). Both means suggest a positive but small relationship.

Studies using panel data and Bloomberg's ESG data yield more conservative estimates. In contrast, the raw subgroup means are somewhat higher among studies that omit a corporate social responsibility committee control (0.30 versus 0.23), while adequacy of endogeneity control makes little difference (0.28 versus 0.26). Section~\ref{chap:four} shows that, of the patterns just described, only the panel-data one survives once other study characteristics are accounted for. The effect also seems more pronounced in the Middle East than in Europe or the United States. Publication status matters as well: published studies report larger effects than unpublished ones.

Whether the literature is subject to publication bias remains an open question. The weighted mean of the estimated effect does differ substantially between published and unpublished studies, but simple averages can obscure the underlying distortions in the reported findings \citep{Ioannidis2017}. The next section therefore tests formally for publication bias.

\section{Publication Bias}
\label{chap:three}

Publication bias shows up in a meta-analysis as a correlation between reported estimates and their standard errors: when significant results of the expected sign are more likely to be published, less precise studies must report larger effects to clear the bar for significance \citep{Stanley2005}. We look for this pattern first visually, with a funnel plot, and then test for it formally.

\begin{figure}[!t]
    \centering
    \caption{Funnel plot of reported estimates}
    \includegraphics[width=\textwidth]{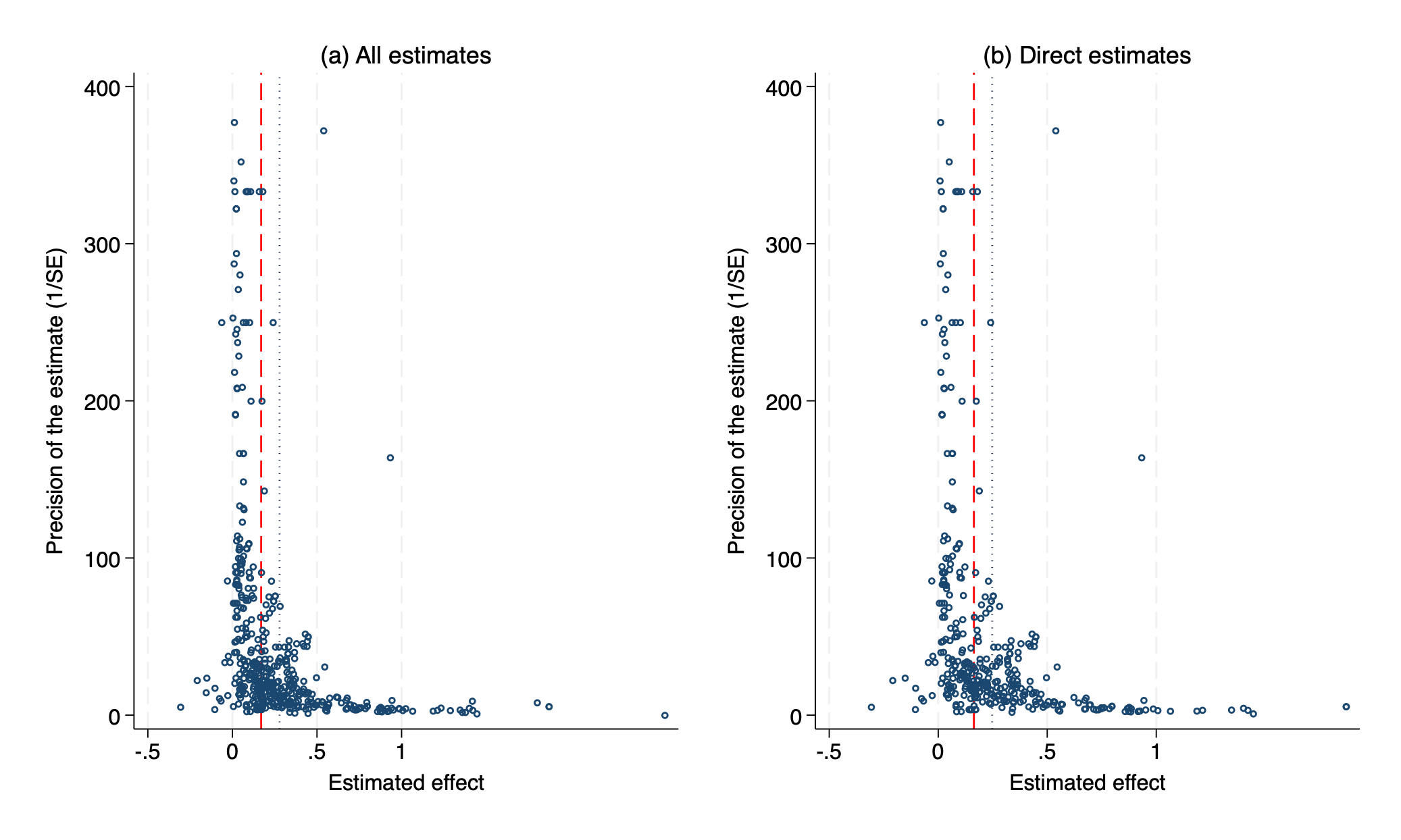}
    \begin{threeparttable}
        \begin{tablenotes}
            \footnotesize
            \item \textit{Notes:} Panel~(a) shows all estimates; panel~(b) restricts to direct estimates. In the absence of publication bias, the most precise estimates 
            are expected to cluster around the mean effect. Less precise estimates should be symmetrically distributed around it. The figure suggests an asymmetry, with the most 
            precise estimates located between zero and the mean effect (dotted vertical line). 
            The dashed line represents the median effect. Extreme outliers are omitted from the 
            figure but remain included in all statistical tests.
        \end{tablenotes}
    \end{threeparttable}
    \label{fig:funnel_full}
\end{figure}

The funnel plot \citep{Egger1997} shows individual effect estimates on the horizontal axis against their precision, the inverse standard error, on the vertical axis. If no publication bias is present, the most precise estimates cluster at the top of the graph around the average effect, and the spread of points widens toward the bottom, forming a symmetric inverted funnel \citep{Stanley2005}. The reason is that only random sampling variation then separates the reported findings from the true effect; less precise estimates stray farther, but in no preferred direction \citep{Sterne2004}.

Applying a funnel plot to our sample of estimates produces the slightly skewed inverted funnel depicted in Figure \ref{fig:funnel_full}. The most precise estimates cluster to the left of the mean effect, consistent with publication bias. The funnel plot for direct, non-imputed estimates shows the same asymmetry: negative estimates are slightly underrepresented, and the most precise estimates lie between zero and the sample mean of the estimated effect.  Funnel asymmetry has other possible sources, though: heterogeneity between studies, differences in methodology, or data quality issues \citep{Egger1997}. The plot only suggests asymmetry; we test for it formally below.

\citet{Egger1997} propose a linear test of this asymmetry, known as the Egger regression, which regresses the reported estimates on their standard errors. If publication bias is present, the correlation will be significantly different from zero \citep{Stanley2005}.

Following the now-standard calibration of \citet{Doucouliagos2013} for gauging the severity of publication selection, if the funnel asymmetry test (FAT) is not statistically significant or the absolute value of \(\beta_{1}\), the coefficient on the standard error in the Egger regression, is less than 1, the degree of publication bias is classified as little to modest. Publication bias is considered substantial if FAT is significant and the absolute value of \(\beta_{1}\) ranges between 1 and 2. If \(\beta_{1}\) exceeds 2 and FAT is significant, publication bias is classified as severe.

We estimate the Egger regression in four specifications. First, we use ordinary least squares. Second, we estimate a model that uses the between-study variance. A model that employs within-study variance is not estimated, as some primary studies provide only one effect estimate. Third and fourth, we employ two weighting schemes commonly applied in meta-analyses, following \citet{Gechert2022} and \citet{Havranek2018}: weighting by the inverse of the number of estimates per study and by precision of estimates. The first weighting gives each study equal weight regardless of how many estimates it reports; the second gives less weight to less precise estimates and removes the heteroskedasticity built into the FAT equation.

We address the potential heteroskedasticity in the FAT-PET framework by clustering standard errors at the study level. The standard assumption of independently and identically distributed error terms is likely violated because of within-study correlations among reported estimates. Clustering at the study level accounts for this dependence while still assuming independence across studies. Although the number of clusters in our sample is well above the usual minimum for valid inference, the presence of unequal cluster sizes may still introduce bias, as noted by \citet{Mackinnon2017}. To mitigate this issue, we adopt the wild cluster bootstrap of \citet{roodman2019fast}, which is particularly robust to cluster imbalance. In our data the imbalance is moderate: the number of estimates per study ranges from one to 28 (median two), and no single study contributes more than 5.3\% of the sample, which limits the influence any one cluster can exert on the bootstrap inference. We report 95\% confidence intervals based on this procedure for all specifications, with the exception of the between-effects estimation at the study level.

Finally, because estimating the Egger regression might suffer from endogeneity between the estimates and their standard errors, we follow \citet{Irsova2024} and instrument the reported variance using the meta-analysis instrumental variable estimator (MAIVE). This endogeneity usually arises from random sampling errors and the joint computation of estimates and their standard errors, both potentially influenced by the estimation method chosen in the primary study. The instrument is the inverse of the number of observations: studies with larger samples should produce smaller standard errors (relevance), and sample size should not be correlated with the chosen estimation method (exogeneity). 

Panel A of Block 1 in Table \ref{tab:pubbias_main} summarizes the results from the model specifications described above. All four regression-based specifications show positive, substantial to severe publication bias significant at the 1\% level; MAIVE does not identify a separate bias coefficient and its corrected mean is statistically indistinguishable from zero given the weak instrument. Compared to the weighted and unweighted means from the primary studies (0.278 and 0.269, see Table \ref{tab:descr_stat}), the bias-corrected mean is notably lower, ranging from 0.076 to 0.113. MAIVE pulls the corrected mean essentially all the way to zero; however, given the reported F-statistics, the instrument is weak, so its point estimate remains unreliable.\footnote{The figures in curly brackets are \citet{anderson1949estimation} 95\% confidence intervals, which are robust to weak identification and are obtained by inverting the Anderson--Rubin test rather than from the point estimate's standard error. They therefore need not be centered on, or even contain, the MAIVE point estimate; in the full-sample and direct-estimate blocks the interval lies entirely above it. We report these intervals for completeness but, given the weak first stage, do not base any conclusion on MAIVE's point estimate.}

Funnel asymmetry and precision-effect tests (FAT-PET) are widely used to detect publication bias and to estimate the true underlying effect \citep{Stanley2008, StanleyDoucouliagos2014}, but they assume a linear relationship between standard errors and estimates. This assumption may not hold in practice, especially for highly precise estimates: their inherently small standard errors can yield significance even without selective reporting \citep{Stanley2010}, so such estimates are less likely to be influenced by publication bias. Linear models may therefore exaggerate the extent of bias and underestimate the true effect. We complement the linear FAT-PET approach with a set of non-linear estimators. 

As a first step in our non-linear checks, we implement the weighted average of adequately powered (WAAP) estimator, introduced by \citet{Stanley2017}. The estimator retains only the estimates whose statistical power to detect the (unrestricted) weighted-average effect exceeds 80\%; the less precise, underpowered estimates that could inflate bias drop out. The adequately powered estimates are then aggregated using optimal inverse-variance weights (\(1/SE^2\)) to derive a more reliable average effect size. However, a known limitation of \acs{waap}, as noted by \citet{Stanley2017}, is its dependency on the presence of sufficiently powered studies within the dataset. When such studies are scarce or absent, \acs{waap} becomes uninformative. A further caveat is that the power screen is benchmarked against the unrestricted weighted-average effect, which is itself not immune to the selection we are trying to correct; we therefore treat \acs{waap} as one input among several rather than as a stand-alone correction.

\begin{table}[!b]
\centering
\begin{minipage}{0.95\linewidth}
\centering
\caption{Tests for publication bias: full sample and direct estimates}
\label{tab:pubbias_main}
\vspace{0.2cm}
\resizebox{\linewidth}{!}{%
\begin{tabular}{l*{5}{c}}
\toprule
\multicolumn{6}{l}{\textbf{Block 1: Full sample}} \\
\midrule
\textit{Panel A: Linear} & OLS & Between Effects & Study Weight & Precision Weight & MAIVE \\
\midrule
Publication Bias            & 1.791\sym{***} & 2.127\sym{***} & 1.837\sym{***} & 2.174\sym{***} & \\
(\textit{Standard Error})   & (0.282)        & (0.210)        & (0.278)        & (0.308)        & \\
                            & [1.175, 2.421] &                & [1.062, 2.505] & [1.474, 2.859] & \\
\addlinespace
Mean Beyond Bias            & 0.113\sym{***} & 0.076\sym{**}  & 0.110\sym{***} & 0.079\sym{***} & $-$0.017 \\
(\textit{Constant})         & (0.020)        & (0.031)        & (0.026)        & (0.020)        & (0.098) \\
                            & [0.072, 0.156] &                & [0.057, 0.161] & [0.031, 0.125] & \{0.039, 0.191\} \\
\addlinespace
First-stage robust F-stat   &  &  &  &  & 6.76 \\
\midrule
\textit{Panel B: Non-linear} & WAAP & Selection Model & Stem method & Endogenous Kink & p-uniform* \\
\midrule
Publication Bias            &           & $P=0.270$ &           & 1.806\sym{**} & \\
                            &           & (0.042)   &           & (0.833)       & \\
\addlinespace
Effect Beyond Bias          & 0.087\sym{***} & 0.113\sym{***} & 0.173\sym{***} & 0.084\sym{***} & 0.174\sym{***} \\
                            & (0.006)        & (0.010)        & (0.013)        & (0.004)        & (0.029) \\
\addlinespace
\# of estimates             & 220 &  & 106  &  & \\
\% of information           &  &  & 100\%  &  & \\
\midrule
Observations & 533 & 533 & 533 & 533 & 533 \\
Studies      & 106 & 106 & 106 & 106 & 106 \\
\bottomrule
\end{tabular}%
} 
\end{minipage}
\end{table}

\begin{table}[!t]
\centering
\addtocounter{table}{-1}
\begin{minipage}{0.95\linewidth}
\centering
\caption{Tests for publication bias: full sample and direct estimates (continued)}
\vspace{0.2cm}
\resizebox{\linewidth}{!}{%
\begin{tabular}{l*{5}{c}}
\toprule
\multicolumn{6}{l}{\textbf{Block 2: Direct estimates}} \\
\midrule
\textit{Panel A: Linear} & OLS & Between Effects & Study Weight & Precision Weight & MAIVE \\
\midrule
Publication Bias            & 1.957\sym{***} & 1.721\sym{***} & 2.098\sym{***} & 2.312\sym{***} & \\
(\textit{Standard Error})   & (0.323)        & (0.312)        & (0.294)        & (0.354)        & \\
                            & [1.077, 2.787] &                & [1.236, 3.078] & [1.454, 3.128] & \\
\addlinespace
Mean Beyond Bias            & 0.104\sym{***} & 0.101\sym{***} & 0.098\sym{***} & 0.078\sym{***} & 0.062 \\
(\textit{Constant})         & (0.019)        & (0.033)        & (0.020)        & (0.019)        & (0.046) \\
                            & [0.064, 0.141] &                & [0.059, 0.139] & [0.029, 0.127] & \{0.103, 0.190\} \\
\addlinespace
First-stage robust F-stat   &  &  &  &  & 10.32 \\
\midrule
\textit{Panel B: Non-linear} & WAAP & Selection Model & Stem method & Endogenous Kink & p-uniform* \\
\midrule
Publication Bias            &           & $P=0.284$ &           & 2.043\sym{**} & \\
                            &           & (0.053)   &           & (1.000)       & \\
\addlinespace
Effect Beyond Bias          & 0.085\sym{***} & 0.115\sym{***} & 0.170\sym{***} & 0.081\sym{***} & 0.153\sym{***} \\
                            & (0.007)        & (0.013)        & (0.013)        & (0.005)        & (0.036) \\
\addlinespace
\# of estimates             & 187 &  & 94  &  & \\
\% of information           &  &  & 99.9\%  &  & \\
\midrule
Observations & 426 & 426 & 426 & 426 & 426 \\
Studies      & 95  & 95  & 95  & 95  & 95 \\
\bottomrule
\end{tabular}%
} 

\footnotesize
{\justifying \textit{Notes:} Block~1 is the full sample; Block~2 restricts to direct estimates. \textit{Panel A:} FAT-PET regression
$E_{is} = \beta_0 + \beta_1 \cdot \text{SE}(E_{is}) + \varepsilon_{is}$, where
$\beta_1$ is publication bias and the constant $\beta_0$ is the mean beyond bias. Cluster-robust standard errors are in parentheses, wild-bootstrap CIs in square brackets \citep{roodman2019fast} and the \citet{anderson1949estimation} 95\% CI in curly brackets for MAIVE by \citet{Irsova2024} that corrects for spurious precision. Significance stars follow the wild-bootstrap p-values, except in the Between Effects column, which reports conventional between-effects inference.
\textit{Panel B:} WAAP denotes weighted average of adequately powered estimates \citep{Stanley2017}; Selection model denotes the technique due to \citet{Andrews2019}; Stem denotes the
stem-based technique \citep{Furukawa2019}; Kink denotes the endogenous kink model \citep{Bom2019}; p-uniform* denotes the technique due to \citet{Aert2026}. 
Significance: \sym{*} $p<0.10$, \sym{**} $p<0.05$, \sym{***} $p<0.01$\par}
\end{minipage}
\end{table}

The selection model of \citet{Andrews2019} builds on the assumption that the likelihood of publishing an effect estimate is influenced by its statistical significance. The model posits that this probability shifts once the estimate surpasses certain t-statistic thresholds. Maximum likelihood then delivers a publication probability for each interval of t-statistics the thresholds define, and estimates underrepresented in an interval are weighted up. The approach follows the selection model proposed earlier by \citet{Hedges1992}.

We also apply two non-linear methods, the stem-based method and the endogenous kink method, both of which extend the logic behind the \say{Top 10} approach introduced by \citet{Stanley2010}. These approaches assume that a subsample of the most precise estimates is less prone to publication bias, thus providing a more accurate estimate of the true underlying effect. \citet{Furukawa2019} selects the \say{stem} of the funnel plot, the most precise estimates, by minimizing the mean squared error. \citet{Bom2019} instead fit a piecewise linear meta-regression of estimates on their standard errors: a flat segment where the standard error does not move the estimate, and a positively sloped segment where publication bias ties estimates to their standard errors. The \say{kink} where the two segments meet is the precision threshold. Both methods determine the share of precise estimates endogenously, trading efficiency against contamination from imprecise estimates.

As a final non-linear check that avoids regressing estimates on their standard errors, we employ the p-uniform* method developed by \citet{Aert2026}. In our case, p-uniform* is estimated by maximum likelihood. The method rests on the principle that, at the true effect size, the p-values implied by the reported estimates should follow a uniform distribution; publication bias distorts this distribution because statistically significant estimates are over-represented in the published record. Unlike the original p-uniform, which conditions on statistical significance, p-uniform* uses significant and non-significant estimates alike and jointly estimates the mean effect and the between-study heterogeneity. The effect value at which the implied p-value distribution is uniform is the bias-corrected mean.

Panel B of Block 1 in Table \ref{tab:pubbias_main} reports the results of the non-linear tests. On average, the corrected effects run slightly higher than their linear counterparts, ranging from 0.084 to 0.174.

The primary robustness check reported in Block 2 of Table \ref{tab:pubbias_main} excludes the 107 estimates that were transformed or had their uncertainty statistics imputed. For MAIVE the first-stage F rises to 10.32, at the conventional threshold, and the weak-identification-robust Anderson--Rubin interval (0.103 to 0.190) lies entirely above zero, corroborating a small positive effect. The point estimate itself remains uninformative. We repeat the tests on two additional samples, one excluding Middle East observations and one using raw unwinsorized data, both reported in Table~\ref{tab:pubbias_robustness} in Appendix~B. Excluding Middle East observations closely replicates the full-sample findings. On raw, unwinsorized data the linear evidence of publication bias weakens and some linear corrected means rise, whereas the non-linear estimates remain small and positive (Table~\ref{tab:pubbias_robustness}). The F-statistic falls below the conventional threshold of 10 in both cases, indicating a weak instrument. Finally, because our sample pools two ESG-rating providers, and because ESG scores from different providers may yield different results \citep{Dorfleitner2015, Berg2022}, we test whether the results depend on the provider. An interacted funnel-asymmetry test that allows both the slope and the intercept to differ for Bloomberg-rated estimates finds neither difference statistically significant. We cannot reject provider invariance (Table~\ref{tab:fatpet_provider_int} in Appendix~B).

\section{Heterogeneity}
\label{chap:four}

The correlation between estimates and their standard errors, so far attributed to publication bias, may partly reflect heterogeneity in the literature. We therefore test whether the publication-bias results survive the inclusion of control variables that reflect study design, and identify which of these variables systematically explain heterogeneity among the reported estimates. According to \citet{Adams2015}, these could be the use of different samples, time windows or empirical methods. To explore these sources, we codify 23 study characteristics organized into six categories: data characteristics, publication characteristics, estimation techniques, design of the analysis, control variables, and spatial variation.

\begin{table}[!b]
\centering

\begin{minipage}{0.95\linewidth}
\centering
\caption{Overview and descriptive statistics of contextual variables}
\label{tab:desr_heterogeneity_part1}
\vspace{0.2cm}
\resizebox{\linewidth}{!}{%
\begin{tabular}{lllll}
\toprule
\textbf{Variable} & \textbf{Description} & \textbf{Mean} & \textbf{SD} & \textbf{WM} \\
\midrule

\hspace*{0.5cm}Estimate & = estimated effect & 0.269 & 0.338 & 0.278 \\
\hspace*{0.5cm}Standard error & = estimated standard error of the estimate & 0.087 & 0.124 & 0.095 \\
\midrule

\emph{Data characteristics} \\
\hspace*{0.5cm}Sample size & = logarithm of the sample size (no.\ of observations) & 7.327 & 1.732 & 6.896 \\
\hspace*{0.5cm}Data: panel & = 1 if panel data is used for estimation & 0.961 & 0.195 & 0.931 \\
\hspace*{0.5cm}ESG data: LSEG & = 1 if LSEG's ESG data is employed & 0.617 & 0.487 & 0.623 \\
\hspace*{0.5cm}Average data year & = logarithm of the average data year & 7.609 & 0.001 & 7.609 \\
\hspace*{0.5cm}Average board gender diversity & \begin{tabular}[c]{@{}l@{}}= logarithm of the average sample board\\gender diversity\end{tabular} & 2.795 & 0.641 & 2.798 \\

\midrule

\emph{Publication characteristics} \\
\hspace*{0.5cm}Published  & = 1 if published in a  journal & 0.949 & 0.220 & 0.896 \\
\hspace*{0.5cm}Unpublished & \begin{tabular}[c]{@{}l@{}}= 1 if unpublished working paper\\(reference category for publication status)\end{tabular} & 0.051 & 0.220 & 0.104 \\
\hspace*{0.5cm}Number of citations & = logarithm of total citations & 2.871 & 1.538 & 2.930 \\

\midrule

\emph{Design of the analysis} \\
\hspace*{0.5cm}Endogeneity control: poor & = 1 if poor endogeneity control is employed & 0.816 & 0.388 & 0.784 \\
\hspace*{0.5cm}Endogeneity control: proper & \begin{tabular}[c]{@{}l@{}}= 1 if proper endogeneity control is employed\\(reference category for endogeneity control)\end{tabular} & 0.184 & 0.388 & 0.216 \\
\hspace*{0.5cm}Number of variables & \begin{tabular}[c]{@{}l@{}}= logarithm of the number of variables used\\in the estimation\end{tabular} & 2.311 & 0.466 & 2.162 \\

\midrule

\emph{Control variables} \\
\hspace*{0.5cm}Firm size control& = 1 if study controls for size of the firm & 0.889 & 0.314 & 0.878 \\
\hspace*{0.5cm}Board independence control & = 1 if study controls for independence of the board & 0.670 & 0.471 & 0.682 \\
\hspace*{0.5cm}CSR committee control& \begin{tabular}[c]{@{}l@{}}= 1 if study controls for existence of firm's\\corporate social responsibility committee\end{tabular} & 0.386 & 0.487 & 0.360 \\
\bottomrule
\end{tabular}%
} 
\end{minipage}
\end{table}

\begin{table}[!t]
\centering
\addtocounter{table}{-1}
\begin{minipage}{0.95\linewidth}
\centering
\caption{Overview and descriptive statistics of contextual variables (continued)}
\vspace{0.2cm}
\resizebox{\linewidth}{!}{%
\begin{tabular}{lllll}
\toprule
\textbf{Variable} & \textbf{Description} & \textbf{Mean} & \textbf{SD} & \textbf{WM} \\

\midrule

\emph{Spatial variation}\\
\hspace*{0.5cm}Region: global & = 1 if study employs a global sample of firms & 0.332 & 0.471 & 0.217 \\
\hspace*{0.5cm}Region: Europe & = 1 if study focuses on European firms & 0.341 & 0.475 & 0.371 \\
\hspace*{0.5cm}Region: USA & = 1 if study focuses on US firms & 0.079 & 0.270 & 0.121 \\
\hspace*{0.5cm}Region: Asia & = 1 if study focuses on Asian firms & 0.081 & 0.273 & 0.137 \\
\hspace*{0.5cm}Region: Middle East & \begin{tabular}[c]{@{}l@{}}= 1 if study focuses on firms in the Middle East\end{tabular} & 0.056 & 0.231 & 0.057 \\
\hspace*{0.5cm}Region: other & \begin{tabular}[c]{@{}l@{}}= 1 if study focuses on firms from other\\countries (reference category for region)\end{tabular} & 0.111 & 0.314 & 0.097 \\
\hspace*{0.5cm}Market: emerging & \begin{tabular}[c]{@{}l@{}}= 1 if study focuses on firms in emerging\\markets\end{tabular} & 0.184 & 0.388 & 0.256 \\
\hspace*{0.5cm}Sector: financial & = 1 if study focuses on financial firms & 0.069 & 0.254 & 0.099 \\

\midrule

\emph{Estimation techniques} \\
\hspace*{0.5cm}Method: linear  & = 1 if OLS or GLS is used & 0.349 & 0.477 & 0.376 \\
\hspace*{0.5cm}Method: panel& = 1 if FE or \acs{re} is used & 0.403 & 0.491 & 0.384 \\
\hspace*{0.5cm}Method: IV & = 1 if 2SLS, CF or LIML is used & 0.062 & 0.241 & 0.055 \\
\hspace*{0.5cm}Method: GMM & = 1 if GMM or its extension is used & 0.079 & 0.270 & 0.131 \\
\hspace*{0.5cm}Method: other  & \begin{tabular}[c]{@{}l@{}}= 1 if other estimation technique is used\\(reference category for estimation technique)\end{tabular} & 0.107 & 0.309 & 0.054 \\

\bottomrule
\end{tabular}%
} 

\footnotesize
{\justifying \textit{Notes:}  
SD = standard deviation,  
WM = mean weighted by the inverse of the number of estimates per study,  
LSEG = London Stock Exchange Group,  
CF = control function,  
LIML = limited information maximum likelihood,  
GMM = generalized method of moments,  
CSR = corporate social responsibility.\par}

\end{minipage}
\end{table}

The resulting variables are described in Table~\ref{tab:desr_heterogeneity_part1}. While each could influence the reported board gender diversity effects, only a few are likely to matter consistently.  Adding all variables into one model would likely produce very imprecise estimates, even for the key ones; selecting a single \textit{best} model among all possible combinations would be arbitrary and would ignore the uncertainty that comes with such a decision.

Bayesian model averaging (BMA) treats the model space itself as uncertain. Each combination of explanatory variables forms a potential model and receives a posterior model probability (PMP) reflecting how well it explains the data \citep{raftery1995bayesian}; every model then enters the final estimate with that weight, instead of one being chosen and the rest discarded \citep{eicher2011default, havranek2015heterogeneity}.

For each variable, BMA reports a posterior mean, the average of its coefficient across models weighted by their posterior probabilities; a posterior standard deviation, which adds model uncertainty to sampling uncertainty; and a posterior inclusion probability (PIP), the summed posterior probability of the models in which the variable appears \citep{zeugner2011bayesian}. Following a scale proposed by \citet{kass1995bayes}, the evidence for including a variable may be categorized as weak, positive, strong or decisive, based on the PIP value falling into intervals of 0.5--0.75, 0.75--0.95, 0.95--0.99 and 0.99--1, respectively.

With BMA in place, we estimate the following meta-regression:

\begin{equation}
EE_{is}=\beta_{0}+\beta_{1}SE_{EE_{is}}+\beta_{2}X_{is}+\epsilon_{is}
\label{eq:metaregression}
\end{equation}

\noindent where $EE_{is}$ represents the estimated effect, $X_{is}$ collects the study characteristics from Table~\ref{tab:desr_heterogeneity_part1}, and $SE_{EE_{is}}$ is the standard error of the estimate. $\beta_0$ is a constant, $\beta_1$ captures the direction and intensity of publication bias, and $\epsilon_{is}$ is the error term.

The meta-regression is estimated using the \textit{bms} package in R, which applies the Markov chain Monte Carlo approach via the Metropolis--Hastings algorithm to avoid the infeasibility of computing all $2^{24}$ possible models. Instead of evaluating each model, the sampler concentrates on the models with the highest posterior model probabilities. It proposes the addition, removal, or swap of regressors, accepting each move with probability proportional to the models' relative marginal likelihoods, weighted by the model prior. Posterior inclusion probabilities are then recovered from the resulting sampled model frequencies \citep{zeugner2011bayesian}. To ensure that studies contributing many estimates do not dominate the analysis, each observation is weighted by the inverse of the number of estimates per study, giving each study equal total weight. This is consistent with the weighting applied in the frequentist specifications.

Our baseline BMA estimation adopts the unit information $g$-prior (UIP), which sets the prior to carry the weight of a single observation. \citet{eicher2011default} recommend it as a robust default for BMA given the limited prior information available to us. To address collinearity, we follow \citet{george2010dilution} and apply a dilution prior. When the regressors included in a model are highly correlated, the determinant of their correlation matrix approaches zero, and the dilution prior downweights such models accordingly \citep{hasan2018}. To verify that the sampler reaches its stationary distribution, we report a set of MCMC convergence diagnostics in Appendix C in Table~\ref{tab:diagnostics1} and Figure~\ref{fig:posterior1}.

In addition to the baseline BMA model, we introduce several robustness checks. First, we employ frequentist model averaging (FMA). While BMA incorporates prior beliefs, FMA draws solely from the data at hand, providing a complementary, prior-free check. We follow the meta-analysis implementation of \citet{Havranek2017}, who apply the Mallows model averaging estimator of \citet{hansen2007}. Its weights minimize the Mallows criterion, an unbiased estimator of the model's squared prediction error that balances in-sample fit against a penalty for the effective number of parameters. This technique, inspired by \citet{magnus2010comparison} and refined through orthogonalization of the covariate space \citep{amini2012comparison}, narrows the model space considerably when the moderators are many. We build on the code shared by \citet{havranek2021skilled}. We estimate the FMA over the full set of moderators and report it alongside the baseline BMA in Table~\ref{tab:bma1}; its coefficients closely track the BMA posterior means for the high-inclusion-probability moderators, so the main findings do not hinge on the Bayesian priors. 

Second, we re-estimate the BMA under an alternative prior structure, the Bayesian risk information criterion (BRIC) $g$-prior of \citet{fernandez2001benchmark} paired with a random model prior. Third, we run a BMA that adopts the same model prior and $g$-prior as our baseline BMA, but excludes all Middle East observations. The reason is that the sample board gender diversity is substantially correlated with ``Middle East'' (the full correlation matrix of the moderators is shown in Figure~\ref{fig:correlation} in Appendix~C). Any true dependence of the estimated effect on sample board gender diversity could then be masked by collinearity with the Middle East indicator. The results of both are reported in Appendix C (Table~\ref{tab:bma2}).  

\begin{figure}[!t]
    \centering
    \caption{Model inclusion in Bayesian model averaging}
    \includegraphics[width=\textwidth]{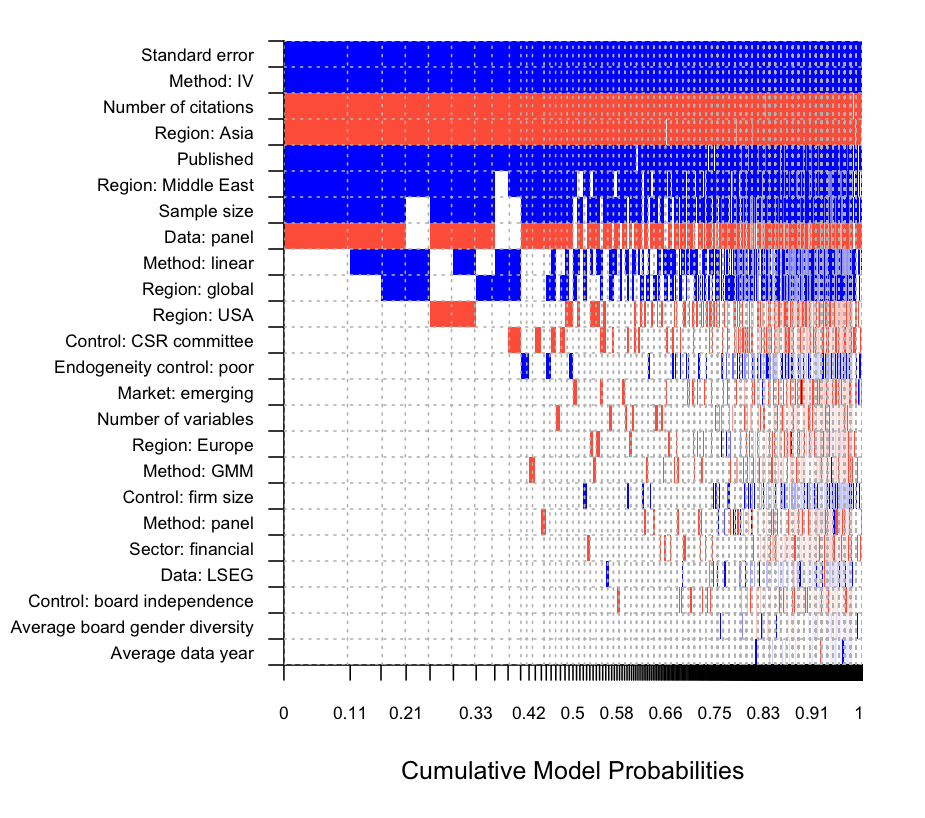}
    \begin{threeparttable}
        \begin{tablenotes}
            \footnotesize
            \item \textit{Notes:} The figure shows results of the baseline BMA estimation
            reported in Table~\ref{tab:bma1} based on the best 7{,}359 models 
            (g-prior = UIP, model prior = dilution). The vertical axis ranks the 
            explanatory variables by their posterior inclusion probabilities from highest 
            (top) to lowest (bottom). The horizontal axis represents the cumulative 
            posterior model probability. A blue (darker) color indicates a positive effect, 
            red (lighter) a negative effect, and absence of color denotes non-inclusion 
            of the corresponding variable in a model. Variable descriptions appear in 
            Table~\ref{tab:desr_heterogeneity_part1}, and further diagnostic details 
            appear in Appendix~C (Table~\ref{tab:diagnostics1}, Figure~\ref{fig:posterior1}).
        \end{tablenotes}
    \end{threeparttable}
    \label{fig:bma1}
\end{figure}

The results of the baseline BMA estimation are visualized in Figure~\ref{fig:bma1}. Variables at the top of the figure are the ones that best explain the estimated effects of board gender diversity on ESG scores, and the models on the left achieve the best trade-off between fit and the number of regressors. A blue cell (darker in grayscale) marks a positive posterior coefficient for the variable in question; a red cell (lighter in grayscale) marks a negative one. The cells for \textit{Middle East}, for example, are blue whenever the variable enters a model: studies of Middle Eastern firms typically report more positive effects. Most of the 24 variables, the figure shows, contribute little to explaining why the reported effects of board gender diversity differ systematically across studies. Only 8 prove robustly important, and each keeps the same sign no matter which other controls enter the model.

\begin{table}[!t]
\centering

\begin{minipage}{0.95\linewidth}
\centering
\caption{Results of baseline BMA and frequentist check estimations}
\vspace{0.2cm}
\label{tab:bma1}

\resizebox{\linewidth}{!}{%
\begin{tabular}{lcccccc}
\toprule
  & \multicolumn{3}{c}{\textbf{Bayesian model averaging}}
  & \multicolumn{3}{c}{\textbf{Frequentist model averaging}} \\
  & \multicolumn{3}{c}{\textbf{(baseline model)}}
  & \multicolumn{3}{c}{\textbf{(frequentist check)}} \\
\cline{2-7}
  & \textit{P. Mean} & \textit{P. SD} & \textit{PIP}
  & \textit{Coef.} & \textit{SE} & \textit{p-value} \\
\midrule

Standard error (SE)            & $\phantom{-}$2.131    & 0.086 & 1.000 & $\phantom{-}$2.085    & 0.095 & 0.000 \\
\midrule

\emph{Data characteristics} \\
\hspace*{0.5cm}Sample size                    & $\phantom{-}$0.030    & 0.015 & 0.840 & $\phantom{-}$0.027    & 0.010 & 0.007 \\
\hspace*{0.5cm}Data: panel                    & $-$0.145              & 0.082 & 0.808 & $-$0.136              & 0.052 & 0.009 \\
\hspace*{0.5cm}ESG data: LSEG                 & $\phantom{-}$0.001    & 0.006 & 0.035 & $\phantom{-}$0.024    & 0.027 & 0.363 \\
\hspace*{0.5cm}Average data year              & $\phantom{-}$0.000    & 0.002 & 0.008 & $\phantom{-}$0.044    & 0.020 & 0.026 \\
\hspace*{0.5cm}Average board gender diversity & $\phantom{-}$0.000    & 0.003 & 0.013 & $-$0.020              & 0.029 & 0.507 \\
\midrule

\emph{Publication characteristics} \\
\hspace*{0.5cm}Published                      & $\phantom{-}$0.156    & 0.050 & 0.965 & $\phantom{-}$0.152    & 0.044 & 0.001 \\
\hspace*{0.5cm}Number of citations            & $-$0.032              & 0.008 & 0.994 & $-$0.034              & 0.007 & 0.000 \\
\midrule

\emph{Design of the analysis} \\
\hspace*{0.5cm}Endogeneity control: poor      & $\phantom{-}$0.005    & 0.017 & 0.106 & $\phantom{-}$0.005    & 0.038 & 0.898 \\
\hspace*{0.5cm}Number of variables            & $-$0.004              & 0.015 & 0.068 & $-$0.062              & 0.031 & 0.049 \\
\midrule

\emph{Control variables} \\
\hspace*{0.5cm}Firm size control              & $\phantom{-}$0.004    & 0.018 & 0.059 & $\phantom{-}$0.090    & 0.040 & 0.024 \\
\hspace*{0.5cm}Board independence control     & $-$0.001              & 0.006 & 0.034 & $-$0.011              & 0.026 & 0.672 \\
\hspace*{0.5cm}CSR committee control          & $-$0.007              & 0.020 & 0.151 & $-$0.036              & 0.027 & 0.180 \\
\midrule

\emph{Spatial variation} \\
\hspace*{0.5cm}Region: global                 & $\phantom{-}$0.036    & 0.051 & 0.395 & $-$0.045              & 0.057 & 0.430 \\
\hspace*{0.5cm}Region: Europe                 & $-$0.003              & 0.015 & 0.064 & $-$0.137              & 0.055 & 0.013 \\
\hspace*{0.5cm}Region: USA                    & $-$0.021              & 0.040 & 0.258 & $-$0.190              & 0.060 & 0.002 \\
\hspace*{0.5cm}Region: Asia                   & $-$0.228              & 0.046 & 0.990 & $-$0.210              & 0.052 & 0.000 \\
\hspace*{0.5cm}Region: Middle East            & $\phantom{-}$0.153    & 0.076 & 0.892 & $\phantom{-}$0.197    & 0.078 & 0.011 \\
\hspace*{0.5cm}Market: emerging               & $-$0.007              & 0.030 & 0.067 & $-$0.171              & 0.060 & 0.004 \\
\hspace*{0.5cm}Sector: financial              & $-$0.001              & 0.010 & 0.047 & $-$0.016              & 0.041 & 0.697 \\
\midrule

\emph{Estimation techniques} \\
\hspace*{0.5cm}Method: linear                 & $\phantom{-}$0.032    & 0.039 & 0.465 & $-$0.012              & 0.053 & 0.822 \\
\hspace*{0.5cm}Method: panel                  & $-$0.002              & 0.011 & 0.061 & $-$0.068              & 0.052 & 0.196 \\
\hspace*{0.5cm}Method: IV                     & $\phantom{-}$0.241    & 0.051 & 1.000 & $\phantom{-}$0.145    & 0.069 & 0.034 \\
\hspace*{0.5cm}Method: GMM                    & $-$0.003              & 0.015 & 0.069 & $-$0.071              & 0.065 & 0.268 \\
\midrule

Studies      & 106 &  &  & 106 &  &  \\
Observations & 533 &  &  & 533 &  &  \\
\bottomrule
\end{tabular}%
} 

\footnotesize
{\justifying \textit{Notes:}  
The table displays results for the baseline BMA model and a frequentist model averaging (FMA) estimation based on the Mallows criterion.  
The FMA is estimated on the full specification and we report its coefficients, standard errors and (two-sided, normal-based) $p$-values for all moderators.
Baseline BMA uses the unit information $g$-prior (UIP) and the dilution prior of \citet{eicher2011default} and \citet{george2010dilution}.
Variables, categories and reference categories are described in Table~\ref{tab:desr_heterogeneity_part1}; the omitted reference categories (unpublished papers, proper endogeneity control, other region, other estimation technique) form the respective baselines.  
P. Mean = Posterior Mean; P. SD = Posterior Standard Deviation; PIP = Posterior Inclusion Probability.\par}

\end{minipage}
\end{table}

Table~\ref{tab:bma1} reports the quantitative counterpart to Figure~\ref{fig:bma1}. In BMA, the posterior mean measures the marginal effect of a study characteristic; the posterior inclusion probability measures how well the characteristic explains the differences among the reported effects. For example, when a study employs panel data the estimated board gender diversity premium typically decreases by almost 0.15 points compared to studies that use cross-sectional data, with a posterior inclusion probability of 81\%. On the 0 to 100 ESG scale, however, a change of 0.15 points is modest. Publication characteristics matter too. Estimated effects from published studies are higher than those from unpublished ones, and a higher number of citations is associated with a lower board gender diversity premium. The estimation method carries the largest posterior mean of any moderator: studies that rely on instrumental-variable techniques report premiums about 0.24 points higher, with a posterior inclusion probability of essentially one. We read this as a methodological rather than a substantive pattern. Instrumental-variable estimates are typically larger and noisier than their OLS or panel counterparts, partly because weak first stages in the primary studies destabilize the estimates and conventional inference. The higher IV premium more plausibly reflects the estimator than a stronger underlying effect.

Yet the largest remaining source of heterogeneity is the geographic location of the firms studied. The baseline and alternative-prior BMA specifications consistently show that studies focusing on Asian firms report a substantially lower board gender diversity premium, while those in the Middle East exhibit a higher one; the check that excludes the Middle East preserves the Asian result. To understand this result, we examine the common characteristics of the underlying observations. In our dataset, the Asian observations are concentrated mainly in a few Southeast Asian emerging markets, predominantly Malaysia, Indonesia and the Philippines. These economies are dominated by family business groups and state-linked conglomerates \citep{claessens2000separation, purdey2016political}. In such firms, female directors are more likely to be members of the controlling family or appointees added to meet board-diversity rules, while the board itself sits beneath a dominant owner. Either way, the marginal woman director is a weak proxy for the independent, stakeholder-oriented governance that is supposed to raise ESG performance. We therefore read this result as a Southeast Asian emerging-market pattern that heavily overlaps with the \textit{Emerging} indicator. It reflects who serves on these boards and how much power they hold; it is not evidence that board gender diversity matters less in Asia as a whole.

In contrast, the studies that cover Middle Eastern countries feature very low average female board representation, typically only a few percent. Such a low base might seem to mechanically amplify the estimated effect of a marginal increase in gender diversity. We favor a more prosaic reading: regional selection or unobserved institutional differences. Such low representation, for instance, often corresponds to tokenism \citep{Kanter2008}, where female board members lack the critical mass necessary to influence firm strategy or governance \citep{kramer2006critical}, particularly in contexts where gender norms may be restrictive. That casts doubt on reading the estimated effect as causal. 

Given concerns about the particular interpretation of the \textit{Middle East} category, we conduct a robustness check on a subsample of the original data, excluding observations from this region; the FMA results already suggest that other regions may drive part of the heterogeneity as well. This check appears in Appendix C: the inclusion probabilities in Table \ref{tab:bma2} reveal no new predictors, and the findings remain robust. Our hypothesis that the effect of sample board gender diversity is masked by collinearity with the Middle East indicator is not supported, as the PIP of sample board gender diversity remains below the 0.5 threshold. This reinforces the interpretation that the larger reported effects for the Middle East may reflect other factors. For example, it is possible that the few firms in the region appointing women to their boards are not representative. They may already excel in ESG performance due to international ownership, progressive values or external pressure. Alternatively, cultural norms or publication incentives may favor the selective reporting of positive effects.  

Two results stand out from the analysis of model uncertainty. First, the evidence of publication bias survives the inclusion of explanatory variables; the standard-error term stays the clearest signal of it, with posterior inclusion probability equal to one and a large positive coefficient. Second, next to standard error, only seven additional variables out of the remaining 23 cross the PIP threshold of 0.5. The rest show no systematic association with the reported effects. 

Based on the moderators from our baseline BMA with posterior inclusion probability above 0.5, we construct a ``best-practice'' estimate using the synthetic study approach. The best-practice estimate is the premium that a well-designed study, free of publication bias, would report. We evaluate it separately for each region (Table~\ref{tab:bpe}) using a dedicated OLS regression on the high-PIP moderators, distinct from the Mallows frequentist model averaging of Table~\ref{tab:bma1}. We set the standard error to zero, assume a published study with panel data, hold sample size, citations, and instrumental-variable use at their means, and vary only the regional indicator. The exact construction is described in the table notes. Setting the standard error to zero treats the entire estimate--standard-error correlation as publication selection. Because such asymmetry can also reflect between-study heterogeneity \citep{Egger1997}, the best-practice figure holds only under that assumption; it is not an assumption-free structural effect.

\begin{table}[!htbp]
\centering
\begin{minipage}{0.80\linewidth}
\centering
\caption{Best-practice estimates of the board gender diversity premium by region}
\label{tab:bpe}
\resizebox{\linewidth}{!}{%
\begin{tabular}{lcc}
\toprule
 & \textit{Best-practice estimate} & \textit{95\% Confidence Interval} \\
\midrule
Rest of world (reference) & $\phantom{-}$0.116 & ($\phantom{-}$0.081 ; $\phantom{-}$0.151) \\
Middle East only              & $\phantom{-}$0.276 & ($\phantom{-}$0.046 ; $\phantom{-}$0.506) \\
Asia only                     & $-$0.108           & ($-$0.212 ; $-$0.003) \\
\bottomrule
\end{tabular}%
}
\vspace{0.15cm}
\footnotesize
{\justifying \textit{Notes:} The table reports best-practice estimates of the board gender diversity premium for three mutually exclusive regions. Each estimate evaluates the moderators with posterior inclusion probability above 0.5 at best-practice values, with the standard error set to zero (free of publication bias), a published study using panel data, with the sample size, number of citations and instrumental-variable use held at their sample means, varying only the regional indicator. Estimates and confidence intervals come from a dedicated OLS regression on the high-PIP moderators (distinct from the Mallows frequentist model averaging reported in Table~\ref{tab:bma1}), weighted by the inverse number of estimates per study and with standard errors clustered at the study level; the interval is the delta-method interval for the linear combination of coefficients, approximated as the estimate $\pm$ 1.96 standard errors.\par}
\end{minipage}
\end{table}

Table~\ref{tab:bpe} reports the resulting estimates. For the rest of the world, the reference region that excludes the Middle East and Asia, the best-practice premium is 0.116 (95\% CI: 0.081 to 0.151). It is small but statistically significant. This is close to, though marginally above, our FAT-PET bias-corrected range of 0.076 to 0.113: once publication bias is removed, the effect a well-designed study would report is modest. On a 0--100 scale, it would be roughly a tenth of an ESG point per one-percentage-point increase in female board representation. The per-percentage-point coefficient, however, understates the practical magnitude of a realistic change. Take a ten-member board, a common size among listed firms: appointing a single additional woman raises board gender diversity by about ten percentage points. Scaling the bias-corrected estimate linearly over such a change implies a gain of roughly one ESG point, against almost three under the raw, uncorrected literature. Because the raw evidence suggests a larger effect where women are scarce and a smaller one where they are already well represented (Figure~\ref{fig:heterogeneity}), this is a rough benchmark rather than a precise prediction. A single point is small against the 0--100 scale, but a change of this order is not negligible for a firm close to a rating threshold. Correction shrinks the premium implied by the raw literature; it does not eliminate it.

But the estimates differ sharply across regions. Studies on Middle Eastern firms imply a best-practice premium of 0.276 (95\% CI: 0.046 to 0.506). It is nearly two and a half times the reference-region estimate and still significantly positive after bias correction. By contrast, studies on Asian firms imply $-0.108$ (95\% CI: $-0.212$ to $-0.003$), below the reference and significantly negative, though the interval only just excludes zero. Geography is thus the dominant source of heterogeneity, with the premium ranging from small and negative in Asia, through a small positive effect in most of the world, to a substantially larger one in the Middle East. As discussed above, neither extreme is best read as a genuine difference in how much board gender diversity matters. The Middle East result most plausibly reflects region-specific selection at very low baseline female board representation, while the Asian result is concentrated in a few Southeast Asian emerging markets dominated by family and state-controlled listed firms. The wide Middle East interval, reflecting the small number of Middle Eastern studies, reinforces this caution. Whether the higher Middle Eastern premium reflects a genuine causal effect or residual region-specific heterogeneity remains an open question.

\section{Conclusion}
\label{chap:five}

Does putting more women on corporate boards reliably improve ESG performance? The question matters for firms facing gender-equity and sustainability targets at the same time. Drawing on 533 estimates from 106 primary studies, we reach a more cautious answer than the published literature suggests. The raw record implies that a one-percentage-point increase in female board representation raises ESG scores by about 0.28 points, on average. A positive association remains after correction, and for a board that adds one woman the implied gain is not trivial. But the association is much smaller than that headline number and concentrated in particular regions, and we cannot rule out that it merely tracks the kinds of firms that appoint women. Correcting for publication bias brings the figure down to between 0.08 and 0.17 points, depending on the method. Our best-practice estimate (a published panel study purged of publication bias) puts the premium at around 0.12 for most of the world, 0.276 for the Middle East, and around $-$0.11 for the Southeast Asian markets that dominate our Asian subsample.

Our analysis yields three main takeaways. First, publication bias is present and robust. The funnel plot is asymmetric and the FAT coefficient stays positive and significant across the main specifications. It survives the inclusion of two dozen control variables and remains of a similar size even when we drop the Middle Eastern studies. Studies with larger standard errors tend to report larger effects, exactly what one would expect when estimates with the intuitive sign and statistical significance are easier to publish. The average reported effect therefore overstates the effect that survives correction. In the meta-regression, the standard-error term remains the strongest single predictor of the reported effects.

Second, the heterogeneity that remains is systematic and mainly driven by geography. Asia enters every BMA specification with a near-unit inclusion probability and a negative sign; the Middle East, wherever the specification includes it, enters with a high inclusion probability and a positive one. We doubt that female directors are simply more effective in the Gulf than in Southeast Asia; the likelier explanation is selection and board composition. In the Middle East, the share of women on boards is very low, often only a few percent, and the few firms that do appoint women may be internationally exposed or otherwise unrepresentative, so the estimated effect would be inflated by this selection rather than by a stronger underlying channel. The Asian observations, by contrast, come mostly from a few Southeast Asian emerging markets dominated by family and state-linked firms, where the women on boards often belong to the controlling family or fill seats created by diversity rules. Both regional results say more about which firms appoint women, and whom they appoint, than about what board gender diversity does.

Third, apart from geography, only a few characteristics systematically shape the reported effects. The estimation method matters most. Studies that rely on instrumental-variable techniques report higher premiums, whereas using panel data lowers the premium by about 0.15 ESG points. Publication status, the number of citations and the sample size also cross the inclusion threshold, while the ESG-rating provider, the control variables a study uses, the development status of the market and the sector of the firms do not.

A few caveats apply. First, our sample only covers studies that use Bloomberg or LSEG's ESG ratings, which limits comparability with research based on other metrics and, indirectly, tends to restrict the analysis to larger, more visible listed companies, since Bloomberg and LSEG ESG coverage concentrates on such firms. Second, the publication-bias methods we use recover a mean effect conditional on the study-design choices we observe in the literature. They cannot recover a structural causal parameter. The regional patterns described above are exactly the kind of residual heterogeneity that such methods are not designed to resolve. Third, our regional best-practice estimates rest on rather thin evidence at the extremes: the Middle Eastern interval is wide because few studies cover the region, and the Asian estimate, though more precisely pinned down, clears conventional significance only marginally.

For policy, board gender diversity remains a goal worth pursuing in its own right, for reasons of fairness, representation, and the quality of governance. Our results do not say that board gender diversity has no effect on ESG. Even after correcting for publication bias, adding one woman to a board of ten is associated, as a rough linear benchmark, with a gain on the order of one ESG point; a firm close to a rating threshold may well find a change of this size relevant. What our results do question is the much larger payoff implied by the raw literature, since that figure is largely a product of publication bias and of regional and methodological differences rather than of a reliable causal effect. The case for mandating board gender diversity only because it is expected to raise ESG scores is therefore weak. The case for pursuing it on its own merits does not rest on ESG scores at all.

\vspace{5mm}

\noindent\textbf{Artificial intelligence use.} The authors collected and coded the data without
AI assistance. Claude Opus 4.8, Claude Fable 5, and Claude Sonnet 5 (Anthropic, through Claude
Code) and GPT-5.6 Sol (OpenAI, through Codex CLI) then assisted in cross-checking the collected
data and the reported numbers, preparing the replication package and the online appendix, and
editing the text. All estimates were
produced by the authors' own Stata and R code and can be reproduced from the public replication
package. The authors are responsible for all of the paper's content.

\vspace{10mm}
\begin{multicols}{2}  
\setlength{\bibsep}{0.0pt}  
\bibliographystyle{abbrvnat}
\bibliography{references.bib}

\begin{thebibliography}{97}
\providecommand{\natexlab}[1]{#1}
\providecommand{\url}[1]{\texttt{#1}}
\expandafter\ifx\csname urlstyle\endcsname\relax
  \providecommand{\doi}[1]{doi: #1}\else
  \providecommand{\doi}{doi: \begingroup \urlstyle{rm}\Url}\fi

\bibitem[Aabo and Giorici(2023)]{Aabo2023}
T.~Aabo and I.~C. Giorici.
\newblock {Do female CEOs matter for ESG scores?}
\newblock \emph{Global Finance Journal}, 56:\penalty0 100722, 2023.

\bibitem[Abdullah et~al.(2024)Abdullah, Yamak, Korzhenitskaya, Rahimi, and
  McClellan]{Abdullah2024}
A.~Abdullah, S.~Yamak, A.~Korzhenitskaya, R.~Rahimi, and J.~McClellan.
\newblock {Sustainable development: The role of sustainability committees in
  achieving ESG targets}.
\newblock \emph{Business Strategy and the Environment}, 33\penalty0
  (3):\penalty0 2250--2268, 2024.

\bibitem[Adams et~al.(2015)Adams, De~Haan, Terjesen, and Van~Ees]{Adams2015}
R.~B. Adams, J.~De~Haan, S.~Terjesen, and H.~Van~Ees.
\newblock {Board diversity: Moving the field forward}.
\newblock \emph{Corporate Governance: An International Review}, 23\penalty0
  (2):\penalty0 77--82, 2015.

\bibitem[AlJanadi(2026)]{aljanadi2025gender}
Y.~AlJanadi.
\newblock Gender diversity and disclosure: a meta-analysis.
\newblock \emph{International Journal of Disclosure and Governance},
  23\penalty0 (1):\penalty0 197--211, 2026.

\bibitem[Alkhawaja et~al.(2023)Alkhawaja, Hu, Johl, and
  Nadarajah]{Alkhawaja2023}
A.~Alkhawaja, F.~Hu, S.~Johl, and S.~Nadarajah.
\newblock {Board gender diversity, quotas, and ESG disclosure: Global
  evidence}.
\newblock \emph{International Review of Financial Analysis}, 90:\penalty0
  102823, 2023.

\bibitem[Amini and Parmeter(2012)]{amini2012comparison}
S.~M. Amini and C.~F. Parmeter.
\newblock {Comparison of model averaging techniques: Assessing growth
  determinants}.
\newblock \emph{Journal of Applied Econometrics}, 27\penalty0 (5):\penalty0
  870--876, 2012.

\bibitem[Anderson and Rubin(1949)]{anderson1949estimation}
T.~W. Anderson and H.~Rubin.
\newblock {Estimation of the parameters of a single equation in a complete
  system of stochastic equations}.
\newblock \emph{The Annals of Mathematical Statistics}, 20\penalty0
  (1):\penalty0 46--63, 1949.

\bibitem[Andrews and Kasy(2019)]{Andrews2019}
I.~Andrews and M.~Kasy.
\newblock {Identification of and correction for publication bias}.
\newblock \emph{American Economic Review}, 109\penalty0 (8):\penalty0
  2766--2794, 2019.

\bibitem[Atif et~al.(2021)Atif, Hossain, Alam, and Goergen]{Atif2021}
M.~Atif, M.~Hossain, M.~S. Alam, and M.~Goergen.
\newblock {Does board gender diversity affect renewable energy consumption?}
\newblock \emph{Journal of Corporate Finance}, 66:\penalty0 101665, 2021.

\bibitem[Bear et~al.(2010)Bear, Rahman, and Post]{Bear2010}
S.~Bear, N.~Rahman, and C.~Post.
\newblock {The impact of board diversity and gender composition on corporate
  social responsibility and firm reputation}.
\newblock \emph{Journal of Business Ethics}, 97:\penalty0 207--221, 2010.

\bibitem[Berg et~al.(2022)Berg, Koelbel, and Rigobon]{Berg2022}
F.~Berg, J.~F. Koelbel, and R.~Rigobon.
\newblock {Aggregate confusion: The divergence of ESG ratings}.
\newblock \emph{Review of Finance}, 26\penalty0 (6):\penalty0 1315--1344, 2022.

\bibitem[Bhatia and Marwaha(2022)]{Bhatia2022}
S.~Bhatia and D.~Marwaha.
\newblock {The influence of board factors and gender diversity on the ESG
  disclosure score: a study on Indian companies}.
\newblock \emph{Global Business Review}, 23\penalty0 (6):\penalty0 1544--1557,
  2022.

\bibitem[Birindelli et~al.(2018)Birindelli, Dell'Atti, Iannuzzi, and
  Savioli]{Birindelli2018}
G.~Birindelli, S.~Dell'Atti, A.~P. Iannuzzi, and M.~Savioli.
\newblock {Composition and activity of the board of directors: Impact on ESG
  performance in the banking system}.
\newblock \emph{Sustainability}, 10\penalty0 (12):\penalty0 4699, 2018.

\bibitem[Bom and Rachinger(2019)]{Bom2019}
P.~R.~D. Bom and H.~Rachinger.
\newblock {A kinked meta-regression model for publication bias correction}.
\newblock \emph{Research Synthesis Methods}, 10\penalty0 (4):\penalty0
  497--514, 2019.

\bibitem[Boulouta(2013)]{Boulouta2013}
I.~Boulouta.
\newblock {Hidden connections: The link between board gender diversity and
  corporate social performance}.
\newblock \emph{Journal of Business Ethics}, 113\penalty0 (2):\penalty0
  185--197, 2013.

\bibitem[Bruna et~al.(2021)Bruna, Dang, Ammari, and Houanti]{Bruna2021}
M.~G. Bruna, R.~Dang, A.~Ammari, and L.~Houanti.
\newblock {The effect of board gender diversity on corporate social
  performance: An instrumental variable quantile regression approach}.
\newblock \emph{Finance Research Letters}, 40:\penalty0 101734, 2021.

\bibitem[Byron and Post(2016)]{ByronPost2016}
K.~Byron and C.~Post.
\newblock {Women on Boards of Directors and Corporate Social Performance: A
  Meta-Analysis}.
\newblock \emph{Corporate Governance: An International Review}, 24\penalty0
  (4):\penalty0 428--442, 2016.

\bibitem[Card and Krueger(1995)]{Card1995}
D.~Card and A.~B. Krueger.
\newblock {Time-series minimum-wage studies: A meta-analysis}.
\newblock \emph{The American Economic Review}, 85\penalty0 (2):\penalty0
  238--243, 1995.

\bibitem[Carrasco et~al.(2015)Carrasco, Francoeur, Labelle, Laffarga, and
  Ruiz-Barbadillo]{Carrasco2015}
A.~Carrasco, C.~Francoeur, R.~Labelle, J.~Laffarga, and E.~Ruiz-Barbadillo.
\newblock {Appointing women to boards: is there a cultural bias?}
\newblock \emph{Journal of Business Ethics}, 129:\penalty0 429--444, 2015.

\bibitem[Cazachevici et~al.(2020)Cazachevici, Havr{\'a}nek, and
  Horv{\'a}th]{Cazachevici2020}
A.~Cazachevici, T.~Havr{\'a}nek, and R.~Horv{\'a}th.
\newblock {Remittances and economic growth: A meta-analysis}.
\newblock \emph{World Development}, 134:\penalty0 105021, 2020.

\bibitem[Chebbi and Ammer(2022)]{Chebbi2022}
K.~Chebbi and M.~A. Ammer.
\newblock {Board composition and ESG disclosure in Saudi Arabia: The moderating
  role of corporate governance reforms}.
\newblock \emph{Sustainability}, 14\penalty0 (19):\penalty0 12173, 2022.

\bibitem[Claessens et~al.(2000)Claessens, Djankov, and
  Lang]{claessens2000separation}
S.~Claessens, S.~Djankov, and L.~H.~P. Lang.
\newblock {The Separation of Ownership and Control in East Asian Corporations}.
\newblock \emph{Journal of Financial Economics}, 58\penalty0 (1-2):\penalty0
  81--112, 2000.

\bibitem[Cook et~al.(2026{\natexlab{a}})Cook, Barto{\v{s}}, Bom, Gechert,
  Kantov{\'a}, Geyer-Klingeberg, Havr{\'a}nek, Ir{\v{s}}ov{\'a}, Luskova,
  Opatrn{\'y}, Prante, Rachinger, and Stanley]{Cook2026ai}
N.~Cook, F.~Barto{\v{s}}, P.~R.~D. Bom, S.~Gechert, K.~Kantov{\'a},
  J.~Geyer-Klingeberg, T.~Havr{\'a}nek, Z.~Ir{\v{s}}ov{\'a}, M.~Luskova,
  M.~Opatrn{\'y}, F.~Prante, H.~J. Rachinger, and T.~D. Stanley.
\newblock {Guidance for the Use of AI in the Meta-Analysis of Economics
  Research}.
\newblock \emph{Journal of Economic Surveys}, 2026{\natexlab{a}}.
\newblock \doi{10.1111/joes.70105}.
\newblock Forthcoming.

\bibitem[Cook et~al.(2026{\natexlab{b}})Cook, Barto{\v{s}}, Bom, Gechert,
  Kantov{\'a}, Geyer-Klingeberg, Havr{\'a}nek, Ir{\v{s}}ov{\'a}, Luskova,
  Opatrn{\'y}, Prante, Rachinger, and Stanley]{Cook2026guidelines}
N.~Cook, F.~Barto{\v{s}}, P.~R.~D. Bom, S.~Gechert, K.~Kantov{\'a},
  J.~Geyer-Klingeberg, T.~Havr{\'a}nek, Z.~Ir{\v{s}}ov{\'a}, M.~Luskova,
  M.~Opatrn{\'y}, F.~Prante, H.~J. Rachinger, and T.~D. Stanley.
\newblock {Reporting Guidelines for Meta-Analysis in Economics---Updated for
  AI}.
\newblock \emph{Journal of Economic Surveys}, 2026{\natexlab{b}}.
\newblock \doi{10.1111/joes.70116}.
\newblock Forthcoming.

\bibitem[Dakhli(2021)]{Dakhli2021}
A.~Dakhli.
\newblock {Does financial performance moderate the relationship between board
  attributes and corporate social responsibility in French firms?}
\newblock \emph{Journal of Global Responsibility}, 12\penalty0 (4):\penalty0
  373--399, 2021.

\bibitem[Dang et~al.(2023{\natexlab{a}})Dang, Hikkerova, Simioni, and
  Sahut]{Dang2023b}
R.~Dang, L.~Hikkerova, M.~Simioni, and J.~M. Sahut.
\newblock {How do women on corporate boards shape corporate social performance?
  Evidence drawn from semiparametric regression}.
\newblock \emph{Annals of Operations Research}, 330\penalty0 (1):\penalty0
  361--388, 2023{\natexlab{a}}.

\bibitem[Dang et~al.(2023{\natexlab{b}})Dang, Houanti, Simioni, and
  Sahut]{Dang2023a}
R.~Dang, L.~Houanti, M.~Simioni, and J.~M. Sahut.
\newblock {The role of endogeneity in the relationship between board gender
  diversity and corporate social performance: evidence from a control function
  method}.
\newblock \emph{Annals of Operations Research}, pages 1--33,
  2023{\natexlab{b}}.

\bibitem[Dorfleitner et~al.(2015)Dorfleitner, Halbritter, and
  Nguyen]{Dorfleitner2015}
G.~Dorfleitner, G.~Halbritter, and M.~Nguyen.
\newblock {Measuring the level and risk of corporate responsibility--An
  empirical comparison of different ESG rating approaches}.
\newblock \emph{Journal of Asset Management}, 16:\penalty0 450--466, 2015.

\bibitem[Doucouliagos and Stanley(2009)]{Doucouliagos2009}
H.~Doucouliagos and T.~D. Stanley.
\newblock {Publication selection bias in minimum-wage research? A
  meta-regression analysis}.
\newblock \emph{British Journal of Industrial Relations}, 47\penalty0
  (2):\penalty0 406--428, 2009.

\bibitem[Doucouliagos and Stanley(2013)]{Doucouliagos2013}
H.~Doucouliagos and T.~D. Stanley.
\newblock {Are all economic facts greatly exaggerated? Theory competition and
  selectivity}.
\newblock \emph{Journal of Economic Surveys}, 27\penalty0 (2):\penalty0
  316--339, 2013.

\bibitem[Egger et~al.(1997)Egger, Smith, Schneider, and Minder]{Egger1997}
M.~Egger, G.~D. Smith, M.~Schneider, and C.~Minder.
\newblock {Bias in meta-analysis detected by a simple, graphical test}.
\newblock \emph{BMJ}, 315\penalty0 (7109):\penalty0 629--634, 1997.

\bibitem[Eicher et~al.(2011)Eicher, Papageorgiou, and
  Raftery]{eicher2011default}
T.~S. Eicher, C.~Papageorgiou, and A.~E. Raftery.
\newblock {Default priors and predictive performance in Bayesian model
  averaging, with application to growth determinants}.
\newblock \emph{Journal of Applied Econometrics}, 26\penalty0 (1):\penalty0
  30--55, 2011.

\bibitem[Fahad and Rahman(2020)]{Fahad2020}
P.~Fahad and P.~M. Rahman.
\newblock {Impact of corporate governance on CSR disclosure}.
\newblock \emph{International Journal of Disclosure and Governance},
  17\penalty0 (2):\penalty0 155--167, 2020.

\bibitem[Fern{\'a}ndez et~al.(2001)Fern{\'a}ndez, Ley, and
  Steel]{fernandez2001benchmark}
C.~Fern{\'a}ndez, E.~Ley, and M.~F.~J. Steel.
\newblock {Benchmark Priors for Bayesian Model Averaging}.
\newblock \emph{Journal of Econometrics}, 100\penalty0 (2):\penalty0 381--427,
  2001.

\bibitem[Fu et~al.(2023)Fu, Wang, and Zhou]{Fu2023}
B.~Fu, K.~Wang, and T.~Zhou.
\newblock {A Controversy in Sustainable Development: How Does Gender Diversity
  Affect the ESG Disclosure?}
\newblock In \emph{{International Conference on Economic Management and Green
  Development}}, pages 669--678. Springer, 2023.

\bibitem[Furukawa(2019)]{Furukawa2019}
C.~Furukawa.
\newblock {Publication bias under aggregation frictions: Theory, evidence, and
  a new correction method}.
\newblock Technical report, Kiel, Hamburg: ZBW--Leibniz Information Centre for
  Economics, 2019.

\bibitem[Gangi et~al.(2021)Gangi, Daniele, Varrone, Vicentini, and
  Coscia]{Gangi2021}
F.~Gangi, L.~M. Daniele, N.~Varrone, F.~Vicentini, and M.~Coscia.
\newblock {Equity mutual funds' interest in the environmental, social and
  governance policies of target firms: Does gender diversity in management
  teams matter?}
\newblock \emph{Corporate Social Responsibility and Environmental Management},
  28\penalty0 (3):\penalty0 1018--1031, 2021.

\bibitem[Gechert et~al.(2022)Gechert, Havr{\'a}nek, Ir\v{s}ov{\'a}, and
  Kolcunov{\'a}]{Gechert2022}
S.~Gechert, T.~Havr{\'a}nek, Z.~Ir\v{s}ov{\'a}, and D.~Kolcunov{\'a}.
\newblock {Measuring capital-labor substitution: The importance of method
  choices and publication bias}.
\newblock \emph{Review of Economic Dynamics}, 45:\penalty0 55--82, 2022.

\bibitem[George(2010)]{george2010dilution}
E.~I. George.
\newblock {Dilution priors: Compensating for model space redundancy}.
\newblock \emph{IMS Collections: Borrowing Strength: Theory Powering
  Applications-A. Festschrift for Lawrence D. Brown}, 6:\penalty0 158--165,
  2010.

\bibitem[Gerged et~al.(2023)Gerged, Tran, and Beddewela]{Gerged2023}
A.~M. Gerged, M.~Tran, and E.~S. Beddewela.
\newblock {Engendering pro-sustainable performance through a multi-layered
  gender diversity criterion: Evidence from the hospitality and tourism
  sector}.
\newblock \emph{Journal of Travel Research}, 62\penalty0 (5):\penalty0
  1047--1076, 2023.

\bibitem[Giannarakis(2013)]{Giannarakis2013}
G.~Giannarakis.
\newblock {Determinants of corporate social responsibility disclosures: the
  case of the US companies}.
\newblock \emph{International Journal of Information Systems and Change
  Management}, 6\penalty0 (3):\penalty0 205--221, 2013.

\bibitem[Gilligan(1977)]{Gilligan1977}
C.~Gilligan.
\newblock {In a different voice: Women's conceptions of self and of morality}.
\newblock \emph{Harvard Educational Review}, 47\penalty0 (4):\penalty0
  481--517, 1977.

\bibitem[Hansen(2007)]{hansen2007}
B.~E. Hansen.
\newblock Least squares model averaging.
\newblock \emph{Econometrica}, 75\penalty0 (4):\penalty0 1175--1189, 2007.

\bibitem[Harjoto et~al.(2015)Harjoto, Laksmana, and Lee]{Harjoto2015}
M.~Harjoto, I.~Laksmana, and R.~Lee.
\newblock {{Board Diversity and Corporate Social Responsibility}}.
\newblock \emph{Journal of Business Ethics}, 132\penalty0 (4):\penalty0
  641--660, 2015.
\newblock \doi{10.1007/s10551-014-2343-0}.

\bibitem[Hasan et~al.(2018)Hasan, Horvath, and Mares]{hasan2018}
I.~Hasan, R.~Horvath, and J.~Mares.
\newblock {What type of finance matters for growth? Bayesian model averaging
  evidence}.
\newblock \emph{The World Bank Economic Review}, 32\penalty0 (2):\penalty0
  383--409, 2018.

\bibitem[Havr{\'a}nek et~al.(2015)Havr{\'a}nek, Horv{\'a}th, Ir\v{s}ov{\'a},
  and Rusn{\'a}k]{havranek2015heterogeneity}
T.~Havr{\'a}nek, R.~Horv{\'a}th, Z.~Ir\v{s}ov{\'a}, and M.~Rusn{\'a}k.
\newblock {Cross-Country Heterogeneity in Intertemporal Substitution}.
\newblock \emph{Journal of International Economics}, 96\penalty0 (1):\penalty0
  100--118, 2015.

\bibitem[Havr{\'a}nek et~al.(2017)Havr{\'a}nek, Rusn{\'a}k, and
  Sokolov{\'a}]{Havranek2017}
T.~Havr{\'a}nek, M.~Rusn{\'a}k, and A.~Sokolov{\'a}.
\newblock {Habit formation in consumption: A meta-analysis}.
\newblock \emph{{European Economic Review}}, 95:\penalty0 142--167, 2017.

\bibitem[Havr{\'a}nek et~al.(2018)Havr{\'a}nek, Ir\v{s}ov{\'a}, and
  Zeynalova]{Havranek2018}
T.~Havr{\'a}nek, Z.~Ir\v{s}ov{\'a}, and O.~Zeynalova.
\newblock {Tuition fees and university enrolment: a meta-regression analysis}.
\newblock \emph{Oxford Bulletin of Economics and Statistics}, 80\penalty0
  (6):\penalty0 1145--1184, 2018.

\bibitem[Havr{\'a}nek et~al.(2020)Havr{\'a}nek, Stanley, Doucouliagos, Bom,
  Geyer-Klingeberg, Iwasaki, Reed, Rost, and van Aert]{Havranek2020guidelines}
T.~Havr{\'a}nek, T.~D. Stanley, H.~Doucouliagos, P.~R.~D. Bom,
  J.~Geyer-Klingeberg, I.~Iwasaki, W.~R. Reed, K.~Rost, and R.~C. van Aert.
\newblock {Reporting Guidelines for Meta-Analysis in Economics}.
\newblock \emph{Journal of Economic Surveys}, 34\penalty0 (3):\penalty0
  469--475, 2020.

\bibitem[Havr{\'a}nek et~al.(2024)Havr{\'a}nek, Ir\v{s}ov{\'a}, Laslopov{\'a},
  and Zeynalova]{havranek2021skilled}
T.~Havr{\'a}nek, Z.~Ir\v{s}ov{\'a}, L.~Laslopov{\'a}, and O.~Zeynalova.
\newblock {Publication and Attenuation Biases in Measuring Skill Substitution}.
\newblock \emph{The Review of Economics and Statistics}, 106\penalty0
  (5):\penalty0 1187--1200, 2024.

\bibitem[Hedges(1992)]{Hedges1992}
L.~V. Hedges.
\newblock {Modeling publication selection effects in meta-analysis}.
\newblock \emph{Statistical Science}, 7\penalty0 (2):\penalty0 246--255, 1992.

\bibitem[Heinemann et~al.(2018)Heinemann, Moessinger, and Yeter]{Heinemann2018}
F.~Heinemann, M.-D. Moessinger, and M.~Yeter.
\newblock {Do fiscal rules constrain fiscal policy? A
  meta-regression-analysis}.
\newblock \emph{European Journal of Political Economy}, 51:\penalty0 69--92,
  2018.

\bibitem[Husted and de~Sousa-Filho(2019)]{Husted2019}
B.~W. Husted and J.~M. de~Sousa-Filho.
\newblock {Board structure and environmental, social, and governance disclosure
  in Latin America}.
\newblock \emph{Journal of Business Research}, 102:\penalty0 220--227, 2019.

\bibitem[Ioannidis et~al.(2017)Ioannidis, Stanley, and
  Doucouliagos]{Ioannidis2017}
J.~P. Ioannidis, T.~D. Stanley, and H.~Doucouliagos.
\newblock {The Power of Bias in Economics Research}.
\newblock \emph{The Economic Journal}, 127\penalty0 (605):\penalty0 F236--F265,
  2017.

\bibitem[Ir\v{s}ov{\'a} et~al.(2024)Ir\v{s}ov{\'a}, Doucouliagos, Havr{\'a}nek,
  and Stanley]{Irsova2023}
Z.~Ir\v{s}ov{\'a}, H.~Doucouliagos, T.~Havr{\'a}nek, and T.~D. Stanley.
\newblock {Meta-analysis of social science research: A practitioner's guide}.
\newblock \emph{Journal of Economic Surveys}, 38\penalty0 (5):\penalty0
  1547--1566, 2024.

\bibitem[Ir\v{s}ov{\'a} et~al.(2025)Ir\v{s}ov{\'a}, Bom, Havr{\'a}nek, and
  Rachinger]{Irsova2024}
Z.~Ir\v{s}ov{\'a}, P.~R.~D. Bom, T.~Havr{\'a}nek, and H.~Rachinger.
\newblock Spurious precision in meta-analysis of observational research.
\newblock \emph{Nature Communications}, 16:\penalty0 8454, 2025.

\bibitem[Issa et~al.(2022)Issa, Zaid, and Hanaysha]{Issa2022}
A.~Issa, M.~A. Zaid, and J.~R. Hanaysha.
\newblock {Exploring the relationship between female director's profile and
  sustainability performance: Evidence from the Middle East}.
\newblock \emph{Managerial and Decision Economics}, 43\penalty0 (6):\penalty0
  1980--2002, 2022.

\bibitem[Kamaludin et~al.(2022)Kamaludin, Ibrahim, Sundarasen, and
  Faizal]{Kamaludin2022}
K.~Kamaludin, I.~Ibrahim, S.~Sundarasen, and O.~Faizal.
\newblock {ESG in the boardroom: evidence from the Malaysian market}.
\newblock \emph{International Journal of Corporate Social Responsibility},
  7\penalty0 (1):\penalty0 4, 2022.

\bibitem[Kamran et~al.(2023)Kamran, Djajadikerta, Mat~Roni, Xiang, and
  Butt]{Kamran2023}
M.~Kamran, H.~G. Djajadikerta, S.~Mat~Roni, E.~Xiang, and P.~Butt.
\newblock {Board gender diversity and corporate social responsibility in an
  international setting}.
\newblock \emph{Journal of Accounting in Emerging Economies}, 13\penalty0
  (2):\penalty0 240--275, 2023.

\bibitem[Kanter(2008)]{Kanter2008}
R.~M. Kanter.
\newblock \emph{{Men and women of the corporation: New edition}}.
\newblock Basic Books, 2008.

\bibitem[Kass and Raftery(1995)]{kass1995bayes}
R.~E. Kass and A.~E. Raftery.
\newblock {Bayes Factors}.
\newblock \emph{Journal of the American Statistical Association}, 90\penalty0
  (430):\penalty0 773--795, 1995.

\bibitem[Khemakhem et~al.(2023)Khemakhem, Arroyo, and
  Montecinos]{Khemakhem2023}
H.~Khemakhem, P.~Arroyo, and J.~Montecinos.
\newblock {Gender diversity on board committees and ESG disclosure: evidence
  from Canada}.
\newblock \emph{Journal of Management and Governance}, 27\penalty0
  (4):\penalty0 1397--1422, 2023.

\bibitem[Kramer et~al.(2006)Kramer, Konrad, and Erkut]{kramer2006critical}
V.~W. Kramer, A.~M. Konrad, and S.~Erkut.
\newblock \emph{Critical mass on corporate boards: Why three or more women
  enhance governance}.
\newblock Wellesley Centers for Women, 2006.

\bibitem[Kravchenko et~al.(2023)Kravchenko, Brezovnik, Mlinaric, and
  Tselinko]{Kravchenko2023}
G.~Kravchenko, B.~Brezovnik, F.~Mlinaric, and I.~Tselinko.
\newblock {Unlocking ESG Potential: The Interaction of Gender Diversity and
  Specialized Skills}.
\newblock \emph{Unpublished}, 2023.

\bibitem[Ley and Steel(2009)]{ley2009effect}
E.~Ley and M.~F. Steel.
\newblock {On the effect of prior assumptions in Bayesian model averaging with
  applications to growth regression}.
\newblock \emph{Journal of Applied Econometrics}, 24\penalty0 (4):\penalty0
  651--674, 2009.

\bibitem[MacKinnon and Webb(2017)]{Mackinnon2017}
J.~G. MacKinnon and M.~D. Webb.
\newblock {Wild bootstrap inference for wildly different cluster sizes}.
\newblock \emph{Journal of Applied Econometrics}, 32\penalty0 (2):\penalty0
  233--254, 2017.

\bibitem[Magnus et~al.(2010)Magnus, Powell, and
  Pr{\"u}fer]{magnus2010comparison}
J.~R. Magnus, O.~Powell, and P.~Pr{\"u}fer.
\newblock {A comparison of two model averaging techniques with an application
  to growth empirics}.
\newblock \emph{Journal of Econometrics}, 154\penalty0 (2):\penalty0 139--153,
  2010.

\bibitem[Manita et~al.(2018)Manita, Bruna, Dang, and Houanti]{Manita2018}
R.~Manita, M.~G. Bruna, R.~Dang, and L.~Houanti.
\newblock {Board gender diversity and ESG disclosure: evidence from the USA}.
\newblock \emph{Journal of Applied Accounting Research}, 19\penalty0
  (2):\penalty0 206--224, 2018.

\bibitem[Mart{\'\i}nez et~al.(2022)Mart{\'\i}nez, Mart{\'\i}n-Cervantes, and
  del Mar Miralles-Quir{\'o}s]{Martinez2022}
M.~d. C.~V. Mart{\'\i}nez, P.~A. Mart{\'\i}n-Cervantes, and M.~del Mar
  Miralles-Quir{\'o}s.
\newblock {Sustainable development and the limits of gender policies on
  corporate boards in Europe. A comparative analysis between developed and
  emerging markets}.
\newblock \emph{European Research on Management and Business Economics},
  28\penalty0 (1):\penalty0 100168, 2022.

\bibitem[Miranda et~al.(2023)Miranda, Delgado, and Branco]{Miranda2023}
B.~Miranda, C.~Delgado, and M.~C. Branco.
\newblock {Board characteristics, social trust and ESG performance in the
  European banking sector}.
\newblock \emph{Journal of Risk and Financial Management}, 16\penalty0
  (4):\penalty0 244, 2023.

\bibitem[Nadeem et~al.(2017)Nadeem, Zaman, and Saleem]{Nadeem2017}
M.~Nadeem, R.~Zaman, and I.~Saleem.
\newblock {Boardroom gender diversity and corporate sustainability practices:
  Evidence from Australian Securities Exchange listed firms}.
\newblock \emph{Journal of Cleaner Production}, 149:\penalty0 874--885, 2017.

\bibitem[Nuhu and Alam(2024)]{Nuhu2024}
Y.~Nuhu and A.~Alam.
\newblock {Board characteristics and ESG disclosure in energy industry:
  evidence from emerging economies}.
\newblock \emph{Journal of Financial Reporting and Accounting}, 22\penalty0
  (1):\penalty0 7--28, 2024.

\bibitem[Page et~al.(2021)Page, McKenzie, Bossuyt, Boutron, Hoffmann, Mulrow,
  Shamseer, Tetzlaff, Akl, Brennan, et~al.]{Page2021}
M.~J. Page, J.~E. McKenzie, P.~M. Bossuyt, I.~Boutron, T.~C. Hoffmann, C.~D.
  Mulrow, L.~Shamseer, J.~M. Tetzlaff, E.~A. Akl, S.~E. Brennan, et~al.
\newblock {The PRISMA 2020 statement: an updated guideline for reporting
  systematic reviews}.
\newblock \emph{BMJ}, 372, 2021.

\bibitem[Post and Byron(2015)]{PostByron2015}
C.~Post and K.~Byron.
\newblock {Women on Boards and Firm Financial Performance: A Meta-Analysis}.
\newblock \emph{Academy of Management Journal}, 58\penalty0 (5):\penalty0
  1546--1571, 2015.

\bibitem[Purdey(2016)]{purdey2016political}
J.~Purdey.
\newblock {Political families in Southeast Asia}.
\newblock \emph{South East Asia Research}, 24\penalty0 (3):\penalty0 319--327,
  2016.

\bibitem[Raftery et~al.(1997)Raftery, Madigan, and
  Hoeting]{raftery1995bayesian}
A.~E. Raftery, D.~Madigan, and J.~A. Hoeting.
\newblock {Bayesian Model Averaging for Linear Regression Models}.
\newblock \emph{Journal of the American Statistical Association}, 92\penalty0
  (437):\penalty0 179--191, 1997.

\bibitem[Roodman et~al.(2019)Roodman, Nielsen, MacKinnon, and
  Webb]{roodman2019fast}
D.~Roodman, M.~{\O}. Nielsen, J.~G. MacKinnon, and M.~D. Webb.
\newblock {Fast and wild: Bootstrap inference in Stata using boottest}.
\newblock \emph{The Stata Journal}, 19\penalty0 (1):\penalty0 4--60, 2019.

\bibitem[Setiani and Novitasari(2024)]{Setiani2024}
E.~P. Setiani and B.~T. Novitasari.
\newblock {Exploring the Impact of Board Attributes on ESG Scores of Indonesian
  Companies}.
\newblock \emph{Nominal Barometer Riset Akuntansi dan Manajemen}, 13\penalty0
  (1):\penalty0 131--143, 2024.

\bibitem[Shahbaz et~al.(2020)Shahbaz, Karaman, Kilic, and Uyar]{Shahbaz2020}
M.~Shahbaz, A.~S. Karaman, M.~Kilic, and A.~Uyar.
\newblock {Board attributes, CSR engagement, and corporate performance: what is
  the nexus in the energy sector?}
\newblock \emph{Energy Policy}, 143:\penalty0 111582, 2020.

\bibitem[Shakil et~al.(2021)Shakil, Tasnia, and Mostafiz]{Shakil2021}
M.~H. Shakil, M.~Tasnia, and M.~I. Mostafiz.
\newblock {Board gender diversity and environmental, social and governance
  performance of US banks: Moderating role of environmental, social and
  corporate governance controversies}.
\newblock \emph{International Journal of Bank Marketing}, 39\penalty0
  (4):\penalty0 661--677, 2021.

\bibitem[Stanley(2001)]{Stanley2001}
T.~D. Stanley.
\newblock {Wheat from chaff: Meta-analysis as quantitative literature review}.
\newblock \emph{Journal of Economic Perspectives}, 15\penalty0 (3):\penalty0
  131--150, 2001.

\bibitem[Stanley(2005)]{Stanley2005}
T.~D. Stanley.
\newblock {Beyond publication bias}.
\newblock \emph{Journal of Economic Surveys}, 19\penalty0 (3):\penalty0
  309--345, 2005.

\bibitem[Stanley(2008)]{Stanley2008}
T.~D. Stanley.
\newblock {Meta-regression methods for detecting and estimating empirical
  effects in the presence of publication selection}.
\newblock \emph{Oxford Bulletin of Economics and Statistics}, 70\penalty0
  (1):\penalty0 103--127, 2008.

\bibitem[Stanley and Doucouliagos(2014)]{StanleyDoucouliagos2014}
T.~D. Stanley and H.~Doucouliagos.
\newblock {Meta-regression approximations to reduce publication selection
  bias}.
\newblock \emph{Research Synthesis Methods}, 5\penalty0 (1):\penalty0 60--78,
  2014.

\bibitem[Stanley et~al.(2010)Stanley, Jarrell, and Doucouliagos]{Stanley2010}
T.~D. Stanley, S.~B. Jarrell, and H.~Doucouliagos.
\newblock {Could it be better to discard 90\% of the data? A statistical
  paradox}.
\newblock \emph{The American Statistician}, 64\penalty0 (1):\penalty0 70--77,
  2010.

\bibitem[Stanley et~al.(2013)Stanley, Doucouliagos, Giles, Heckemeyer,
  Johnston, Laroche, Nelson, Paldam, Poot, Pugh, Rosenberger, and
  Rost]{Stanley2013guidelines}
T.~D. Stanley, H.~Doucouliagos, M.~Giles, J.~H. Heckemeyer, R.~J. Johnston,
  P.~Laroche, J.~P. Nelson, M.~Paldam, J.~Poot, G.~Pugh, R.~S. Rosenberger, and
  K.~Rost.
\newblock {Meta-analysis of economics research reporting guidelines}.
\newblock \emph{Journal of Economic Surveys}, 27\penalty0 (2):\penalty0
  390--394, 2013.

\bibitem[Stanley et~al.(2017)Stanley, Doucouliagos, and Ioannidis]{Stanley2017}
T.~D. Stanley, H.~Doucouliagos, and J.~P. Ioannidis.
\newblock {Finding the power to reduce publication bias}.
\newblock \emph{Statistics in Medicine}, 36\penalty0 (10):\penalty0 1580--1598,
  2017.

\bibitem[Sterne and Harbord(2004)]{Sterne2004}
J.~A. Sterne and R.~M. Harbord.
\newblock {Funnel plots in meta-analysis}.
\newblock \emph{The Stata Journal}, 4\penalty0 (2):\penalty0 127--141, 2004.

\bibitem[Uyar et~al.(2021)Uyar, Kuzey, Kilic, and Karaman]{Uyar2021}
A.~Uyar, C.~Kuzey, M.~Kilic, and A.~S. Karaman.
\newblock {Board structure, financial performance, corporate social
  responsibility performance, CSR committee, and CEO duality: Disentangling the
  connection in healthcare}.
\newblock \emph{Corporate Social Responsibility and Environmental Management},
  28\penalty0 (6):\penalty0 1730--1748, 2021.

\bibitem[van Aert and van Assen(2026)]{Aert2026}
R.~C. van Aert and M.~A. van Assen.
\newblock {Correcting for publication bias in a meta-analysis with the
  p-uniform* method}.
\newblock \emph{Psychonomic Bulletin \& Review}, 33\penalty0 (3):\penalty0 102,
  2026.

\bibitem[Wang et~al.(2022)Wang, Yekini, Babajide, and Kessy]{Wang2022}
Y.~Wang, K.~Yekini, B.~Babajide, and M.~Kessy.
\newblock {Antecedents of corporate social responsibility disclosure: evidence
  from the UK extractive and retail sector}.
\newblock \emph{International Journal of Accounting \& Information Management},
  30\penalty0 (2):\penalty0 161--188, 2022.

\bibitem[Waterstraat et~al.(2021)Waterstraat, Kustner, and
  Koch]{Waterstraat2021}
S.~Waterstraat, C.~Kustner, and M.~Koch.
\newblock {Does board composition taking account of sustainability expertise
  influence ESG ratings? An exploratory study of European banks}.
\newblock In \emph{{Society 5.0: First International Conference, Society 5.0
  2021, Virtual Event, June 22--24, 2021, Revised Selected Papers 1}}, pages
  129--138. Springer, 2021.

\bibitem[Williams(2003)]{Williams2003}
R.~J. Williams.
\newblock {Women on corporate boards of directors and their influence on
  corporate philanthropy}.
\newblock \emph{Journal of Business Ethics}, 42:\penalty0 1--10, 2003.

\bibitem[Wu et~al.(2022)Wu, Furuoka, and Lau]{wu2022corporate}
Q.~Wu, F.~Furuoka, and S.~C. Lau.
\newblock Corporate social responsibility and board gender diversity: a
  meta-analysis.
\newblock \emph{Management Research Review}, 45\penalty0 (7):\penalty0
  956--983, 2022.

\bibitem[Yarram and Adapa(2021)]{Yarram2021}
S.~R. Yarram and S.~Adapa.
\newblock {Board gender diversity and corporate social responsibility: Is there
  a case for critical mass?}
\newblock \emph{Journal of Cleaner Production}, 278:\penalty0 123319, 2021.

\bibitem[Zeugner(2011)]{zeugner2011bayesian}
S.~Zeugner.
\newblock {Bayesian model averaging with BMS}.
\newblock \emph{Tutorial to the R-package BMS}, 2011.

\bibitem[{\v{Z}}igraiov{\'a} and Havr{\'a}nek(2016)]{Zigraiova2016}
D.~{\v{Z}}igraiov{\'a} and T.~Havr{\'a}nek.
\newblock {Bank competition and financial stability: Much ado about nothing?}
\newblock \emph{Journal of Economic Surveys}, 30\penalty0 (5):\penalty0
  944--981, 2016.

\end{thebibliography}
\end{multicols}

\clearpage
\pagenumbering{Roman}
\setcounter{page}{1}

\section*{Appendix A}
\vspace{15mm}

\counterwithin{figure}{section}
\renewcommand{\thefigure}{A.\arabic{figure}}

\setcounter{figure}{0}
\setcounter{table}{0}

\savebox{\prismabox}{%
\begin{tikzpicture}[>=latex, font=\small,
  phase/.style={rectangle, rounded corners=5pt, draw=black, fill=white,
                minimum width=2.4cm, minimum height=1.6cm, text centered,
                font=\bfseries\small},
  box/.style={rectangle, draw=black, fill=white, minimum width=2.8cm,
              text width=2.5cm, minimum height=2.0cm, align=center,
              inner sep=4pt, font=\footnotesize}
  ]
  \node[phase] at (0, 12.0) {Identification};
  \node[phase] at (0,  9.0) {Screening};
  \node[phase] at (0,  6.0) {Eligibility};
  \node[phase] at (0,  3.0) {Included};
  \node[box] (gsid)  at (3.4, 12.0) {Studies identified through Google Scholar I. ($n = 22{,}800$)};
  \node[box] (gsscr) at (3.4, 9.0)  {Studies screened based on the order in Google Scholar ($n = 500$)};
  \node[box] (gsexcl)at (6.8, 9.0)  {Studies excluded based on the abstract and outlet ($n = 208$)};
  \node[box] (gsel)  at (3.4, 6.0)  {Studies assessed in detail for eligibility ($n = 292$)};
  \node[box] (gselx) at (6.8, 6.0)  {Studies excluded due to lack of data or correspondence ($n = 42$) and different metrics ($n = 167$)};
  \node[box] (gsinc) at (3.4, 3.0)  {Studies included based on search ($n = 83$)};
  \node[box] (snid)  at (10.2, 12.0) {Studies identified through snowballing ($n = 591$)};
  \node[box] (snscr) at (10.2, 9.0)  {Studies screened based on number of citations ($n = 100$)};
  \node[box] (snexcl)at (13.6, 9.0)  {Studies excluded based on the abstract and outlet ($n = 65$)};
  \node[box] (snel)  at (10.2, 6.0)  {Studies assessed in detail for eligibility ($n = 35$)};
  \node[box] (snelx) at (13.6, 6.0)  {Studies excluded due to lack of data or correspondence ($n = 20$)};
  \node[box] (sninc) at (10.2, 3.0)  {Studies included based on snowballing ($n = 15$)};
  \node[box] (g2id)  at (17.0, 12.0) {Studies identified through Google Scholar II. ($n = 1{,}740$)};
  \node[box] (g2scr) at (17.0, 9.0)  {Studies screened based on the order in Google Scholar ($n = 50$)};
  \node[box] (g2excl)at (20.4, 9.0)  {Studies excluded based on the abstract and outlet ($n = 32$)};
  \node[box] (g2el)  at (17.0, 6.0)  {Studies assessed in detail for eligibility ($n = 18$)};
  \node[box] (g2elx) at (20.4, 6.0)  {Studies excluded due to lack of data or correspondence ($n = 10$)};
  \node[box] (g2inc) at (17.0, 3.0)  {Studies included based on search ($n = 8$)};
  \node[box] (meta) at (10.2, 0) {Studies included in the meta-analysis ($n = 106$)};
  \draw[->] (gsid.south)  -- (gsscr.north);
  \draw[->] (gsscr.south) -- (gsel.north);
  \draw[->] (gsel.south)  -- (gsinc.north);
  \draw[->] (gsinc.south) -- (meta.north west);
  \draw[->] (gsscr.east)  -- (gsexcl.west);
  \draw[->] (gsel.east)   -- (gselx.west);
  \draw[->] (snid.south)  -- (snscr.north);
  \draw[->] (snscr.south) -- (snel.north);
  \draw[->] (snel.south)  -- (sninc.north);
  \draw[->] (sninc.south) -- (meta.north);
  \draw[->] (snscr.east)  -- (snexcl.west);
  \draw[->] (snel.east)   -- (snelx.west);
  \draw[->] (g2id.south)  -- (g2scr.north);
  \draw[->] (g2scr.south) -- (g2el.north);
  \draw[->] (g2el.south)  -- (g2inc.north);
  \draw[->] (g2inc.south) -- (meta.north east);
  \draw[->] (g2scr.east)  -- (g2excl.west);
  \draw[->] (g2el.east)   -- (g2elx.west);
\end{tikzpicture}}

\begin{figure}[H]
\centering
\scalebox{0.8}{%
\rotatebox{90}{%
\begin{minipage}{\wd\prismabox}
\centering
\caption{PRISMA flow diagram}
\label{fig:prisma}
\vspace{3mm}

\usebox{\prismabox}

\bigskip
{\footnotesize\textit{Notes:} The figure presents the PRISMA flow diagram for the study-selection procedure \citep{Irsova2023, Page2021, Stanley2013guidelines}. Studies are identified in Google Scholar because it searches the full text of articles rather than just titles, abstracts, and keywords. The following query is used: \textbf{``Board'' AND ``Gender'' AND ``Diversity'' AND ``ESG'' AND (``Disclosure'' OR ``Score'' OR ``Performance'')}. The initial search in May 2024 returned more than 22,800 records, of which we screen the first 500 by relevance ranking. Backward snowballing is then performed by compiling the 100 most frequently cited works among the included studies. The search was closed on December 16, 2024, when we ran a further Google Scholar query restricted to 2024 to capture articles published between May and December. Whenever a study had a plausible chance of containing usable estimates, its full text was read and assessed. The complete dataset is available in the online appendix at \url{https://meta-analysis.cz/esg}.}
\end{minipage}}}
\end{figure}

\renewcommand{\thefigure}{A.\arabic{figure}}
\begin{figure}[!t]
    \centering
     \caption{Variation of estimates within and across countries}
    \includegraphics[width=0.8\textwidth]{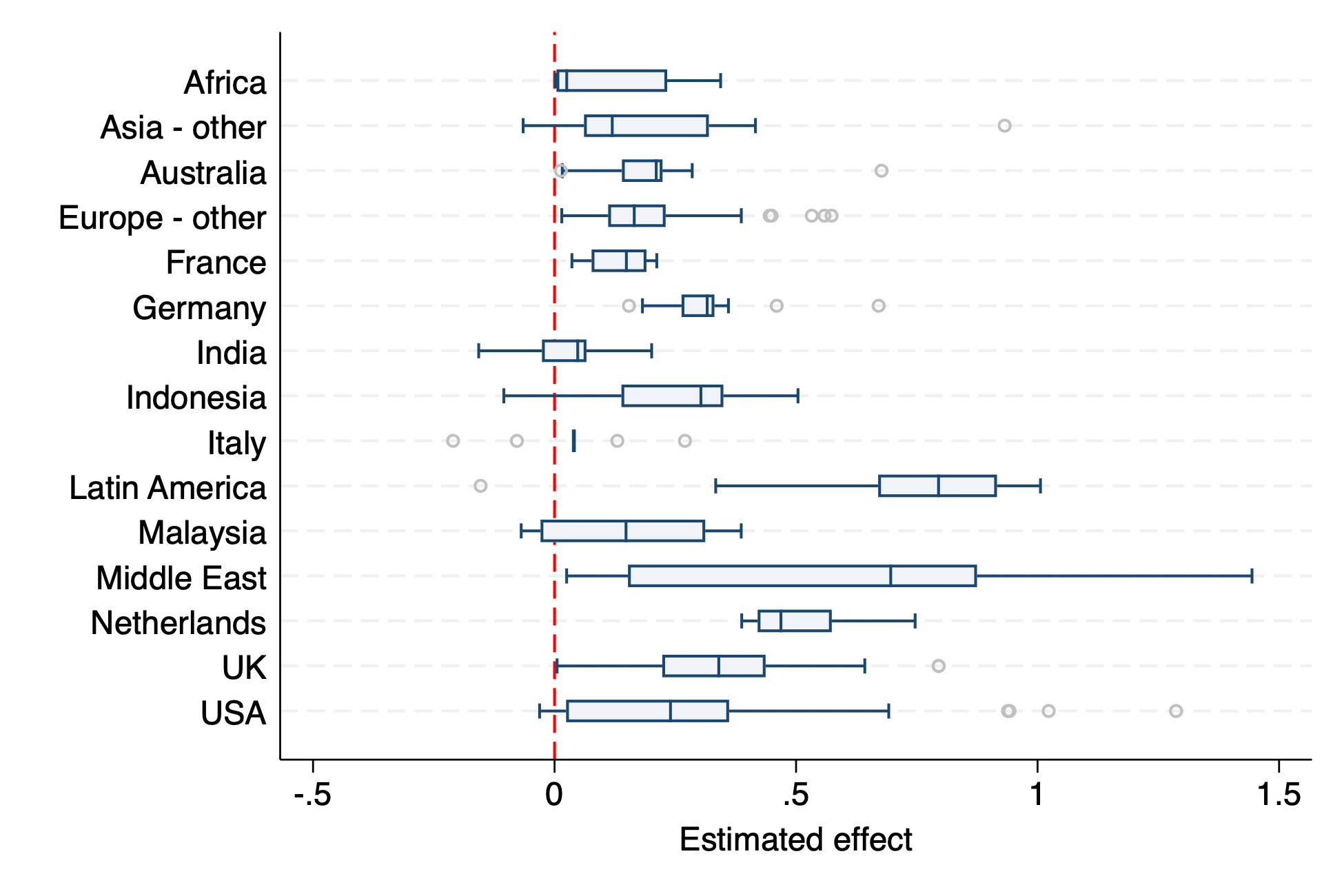}
   \begin{threeparttable}
        \begin{tablenotes}
            \footnotesize
            \item \textit{Notes:} The figure shows a box plot of the estimated effects within the interquartile range. 
            The inner box line indicates the median. The whiskers extend to the most 
            extreme values within 1.5 times the interquartile range. 
        \end{tablenotes}
    \end{threeparttable}
    \label{fig:estimates_hbox}
\end{figure}

\clearpage
\newpage

\section*{Appendix B}

\renewcommand{\thetable}{B.\arabic{table}}
\begin{table}[!b]
\centering
\begin{minipage}{0.9\linewidth}
\centering
\caption{Robustness checks: excluding the Middle East and unwinsorized data}
\label{tab:pubbias_robustness}
\vspace{0.2cm}
\resizebox{\linewidth}{!}{%
\begin{tabular}{l*{5}{c}}
\toprule
\multicolumn{6}{l}{\textbf{Block 1: Subset excluding Middle East}} \\
\midrule
\textit{Panel A: Linear} & OLS & Between Effects & Study Weight & Precision Weight & MAIVE \\
\midrule
Publication Bias            & 1.689\sym{***} & 2.128\sym{***} & 1.628\sym{***} & 2.157\sym{***} & \\
(\textit{Standard Error})   & (0.325)        & (0.218)        & (0.347)        & (0.324)        & \\
                            & [0.853, 2.413] &                & [0.622, 2.328] & [1.410, 2.859] & \\
\addlinespace
Mean Beyond Bias            & 0.116\sym{***} & 0.071\sym{**}  & 0.120\sym{***} & 0.080\sym{***} & $-$0.062 \\
(\textit{Constant})         & (0.022)        & (0.031)        & (0.029)        & (0.020)        & (0.168) \\
                            & [0.072, 0.162] &                & [0.055, 0.183] & [0.033, 0.125] & \{$-$0.239, 0.190\} \\
\addlinespace
First-stage robust F-stat   &  &  &  &  & 3.13 \\
\midrule
\textit{Panel B: Non-linear} & WAAP & Selection Model & Stem method & Endogenous Kink & p-uniform* \\
\midrule
Publication Bias            &           & $P=0.281$ &           & 1.794\sym{**} & \\
                            &           & (0.046)   &           & (0.887)       & \\
\addlinespace
Effect Beyond Bias          & 0.087\sym{***} & 0.114\sym{***} & 0.166\sym{***} & 0.084\sym{***} & 0.167\sym{***} \\
                            & (0.006)        & (0.010)        & (0.013)        & (0.004)        & (0.027) \\
\addlinespace
\# of estimates             & 218 &  & 100  &  & \\
\% of information           &  &  & 100\%  &  & \\
\midrule
Observations & 503 & 503 & 503 & 503 & 503 \\
Studies      & 100 & 100 & 100 & 100 & 100 \\
\addlinespace
\midrule
\multicolumn{6}{l}{\textbf{Block 2: Unwinsorized data}} \\
\midrule
\textit{Panel A: Linear} & OLS & Between Effects & Study Weight & Precision Weight & MAIVE \\
\midrule
Publication Bias            & 0.239            & 0.376\sym{***} & 0.196            & 1.527\sym{***} & \\
(\textit{Standard Error})   & (0.117)          & (0.054)        & (0.074)          & (0.448)        & \\
                            & [$-$9.502, 2.109]  &                & [$-$9.880, 1.940]  & [0.554, 2.620] & \\
\addlinespace
Mean Beyond Bias            & 0.248\sym{***}   & 0.246\sym{***} & 0.307\sym{***}   & 0.083\sym{***} & 0.089 \\
(\textit{Constant})         & (0.032)          & (0.053)        & (0.050)          & (0.019)        & (0.067) \\
                            & [0.183, 0.317]   &                & [0.204, 0.417]   & [0.037, 0.125] & \{$-$0.005, 0.367\} \\
\addlinespace
First-stage robust F-stat   &  &  &  &  & 0.97 \\
\midrule
\textit{Panel B: Non-linear} & WAAP & Selection Model & Stem method & Endogenous Kink & p-uniform* \\
\midrule
Publication Bias            &           & $P=0.277$ &           & 3.821\sym{***} & \\
                            &           & (0.042)   &           & (1.103)        & \\
\addlinespace
Effect Beyond Bias          & 0.038\sym{***} & 0.113\sym{***} & 0.172\sym{***} & 0.060\sym{***} & 0.159\sym{***} \\
                            & (0.005)        & (0.010)        & (0.013)        & (0.003)        & (0.038) \\
\addlinespace
\# of estimates             & 178 &  & 103  &  & \\
\% of information           &  &  & 99.9\%  &  & \\
\midrule
Observations & 533 & 533 & 533 & 533 & 533 \\
Studies      & 106 & 106 & 106 & 106 & 106 \\
\bottomrule
\end{tabular}%
} 

\footnotesize
{\justifying \textit{Notes:} Block~1 excludes Middle East estimates; Block~2 uses unwinsorized data. \textit{Panel A:} FAT-PET regression
$E_{is} = \beta_0 + \beta_1 \cdot \text{SE}(E_{is}) + \varepsilon_{is}$, where
$\beta_1$ is publication bias and the constant $\beta_0$ is the mean beyond bias. Cluster-robust standard errors are in parentheses, wild-bootstrap CIs in square brackets \citep{roodman2019fast} and the \citet{anderson1949estimation} 95\% CI in curly brackets for MAIVE by \citet{Irsova2024} that corrects for spurious precision. Significance stars follow the wild-bootstrap p-values, except in the Between Effects column, which reports conventional between-effects inference.
\textit{Panel B:} WAAP denotes weighted average of adequately powered estimates \citep{Stanley2017}; Selection model denotes the technique due to \citet{Andrews2019}; Stem denotes the
stem-based technique \citep{Furukawa2019}; Kink denotes the endogenous kink model \citep{Bom2019}; p-uniform* denotes the technique due to \citet{Aert2026}. 
Significance: \sym{*} $p<0.10$, \sym{**} $p<0.05$, \sym{***} $p<0.01$\par}
\end{minipage}
\end{table}

\renewcommand{\thetable}{B.\arabic{table}}
\begin{table}[!h]
\centering
\begin{minipage}{0.75\linewidth}
\centering
\caption{Provider comparison: interaction FAT-PET test}
\label{tab:fatpet_provider_int}
\vspace{0.2cm}
\begin{tabular}{l*{2}{c}}
\toprule
 & OLS & Precision Weight \\
\midrule
Publication bias ($\beta_1$)           & 1.624\sym{***}  & 2.350\sym{***} \\
                        & (0.304)         & (0.392) \\
\addlinespace
\quad $\times$ Bloomberg ($\beta_3$)  & 0.397    & $-$0.413 \\
\quad \textit{(difference)}                                                & (0.610)         & (0.636) \\
                                                 & [$-$1.206, 1.783] & [$-$2.041, 0.988] \\
\addlinespace
Mean beyond bias ($\beta_0$)           & 0.135\sym{***}  & 0.075\sym{***} \\
                             & (0.026)         & (0.023) \\
\addlinespace
\quad Bloomberg ($\beta_2$)   & $-$0.057          & 0.011 \\
 \quad \textit{(difference)}                                               & (0.042)         & (0.039) \\
                                                 & [$-$0.149, 0.037] & [$-$0.113, 0.086] \\
\midrule
Observations & 533 & 533 \\
Studies      & 106 & 106 \\
\bottomrule
\end{tabular}

\footnotesize
{\justifying \textit{Notes:} Estimates from the interacted FAT-PET regression
$E_{is} = \beta_0 + \beta_1\,\text{SE}(E_{is}) + \beta_2\,\text{Bloomberg}_s
+ \beta_3\,[\text{SE}(E_{is})\times\text{Bloomberg}_s] + \varepsilon_{is}$,
where $\text{Bloomberg}_s=1$ if the rating provider is Bloomberg (baseline = LSEG ESG data provider).
$\beta_1$ and $\beta_0$ are the funnel-asymmetry slope (publication bias) and corrected mean for the
baseline provider; $\beta_3$ tests whether publication bias differs across providers and $\beta_2$
whether the corrected mean differs. Cluster-robust standard errors (clustered by study) in parentheses.
For the difference terms ($\beta_3$, $\beta_2$), 95\% wild-bootstrap confidence intervals
\citep{roodman2019fast} are in square brackets and significance is from wild-bootstrap p-values.
 Significance: \sym{*} $p<0.10$, \sym{**} $p<0.05$, \sym{***} $p<0.01$.\par}
\end{minipage}
\end{table}

\clearpage

\section*{Appendix C}
\thispagestyle{empty}
\vspace{20mm}
\renewcommand{\thetable}{C.\arabic{table}}
\renewcommand{\thefigure}{C.\arabic{figure}}
\setcounter{figure}{0}
\setcounter{table}{0}

\begin{figure}[!h]
    \centering
    \caption{Correlations between potential predictors of heterogeneity}
    \includegraphics[width=\textwidth]{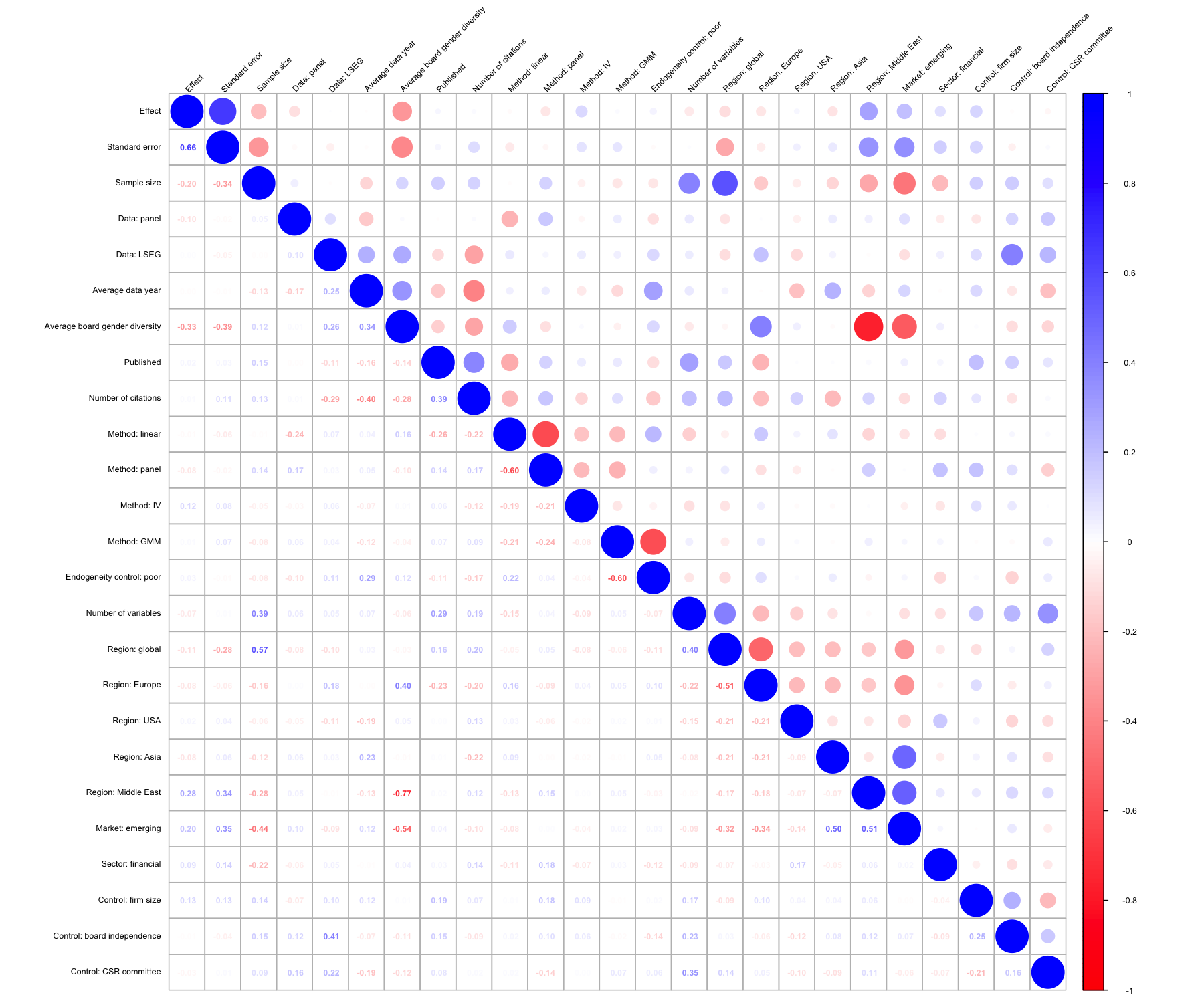}
    \begin{threeparttable}
        \begin{tablenotes}
            \footnotesize
            \item \textit{Notes:} The figure displays correlation coefficients for variables used in heterogeneity analysis.
        \end{tablenotes}
    \end{threeparttable}
    \label{fig:correlation}
\end{figure}

\begin{figure}[!]
    \centering
    \caption{Model size and convergence for the baseline BMA}
    \vspace{0.2cm}
    \begin{minipage}[t]{0.49\textwidth}
        \centering
        \includegraphics[width=\textwidth]{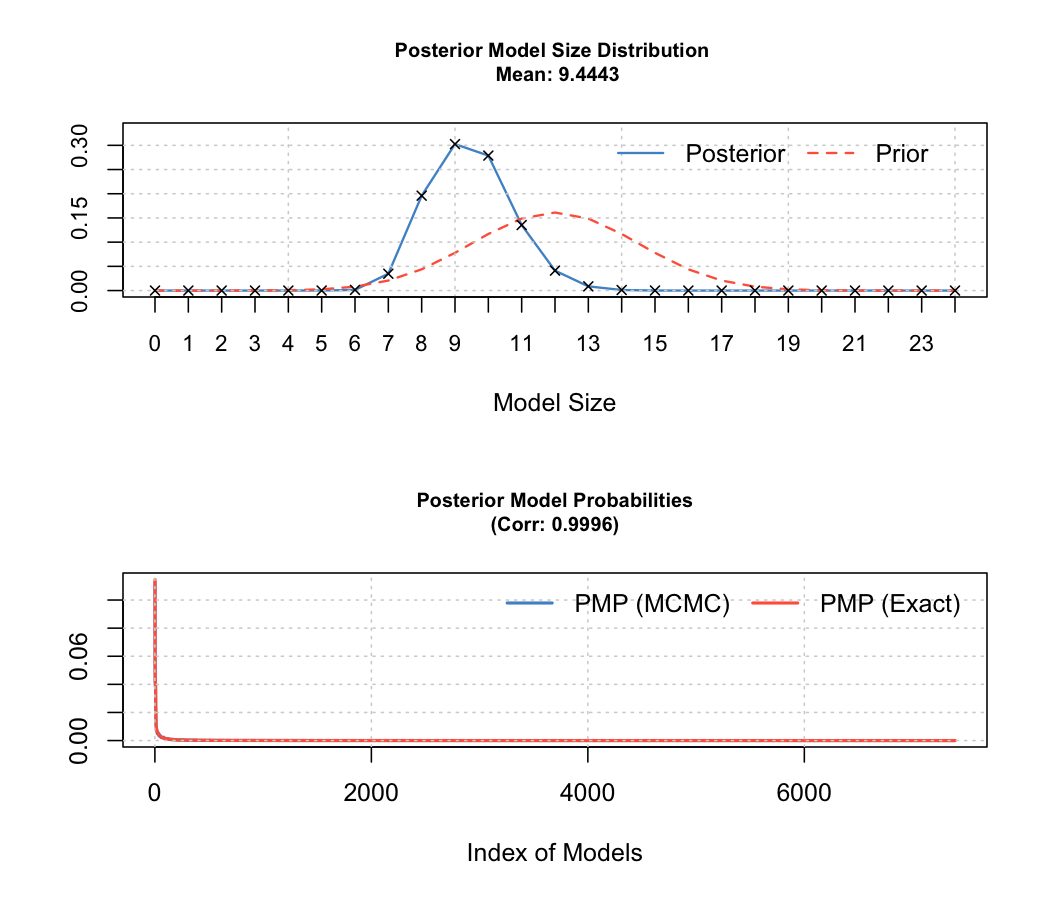}\\[2pt]
        {\footnotesize (a) Posterior model size and model probabilities}
    \end{minipage}\hfill
    \begin{minipage}[t]{0.49\textwidth}
        \centering
        \includegraphics[width=\textwidth]{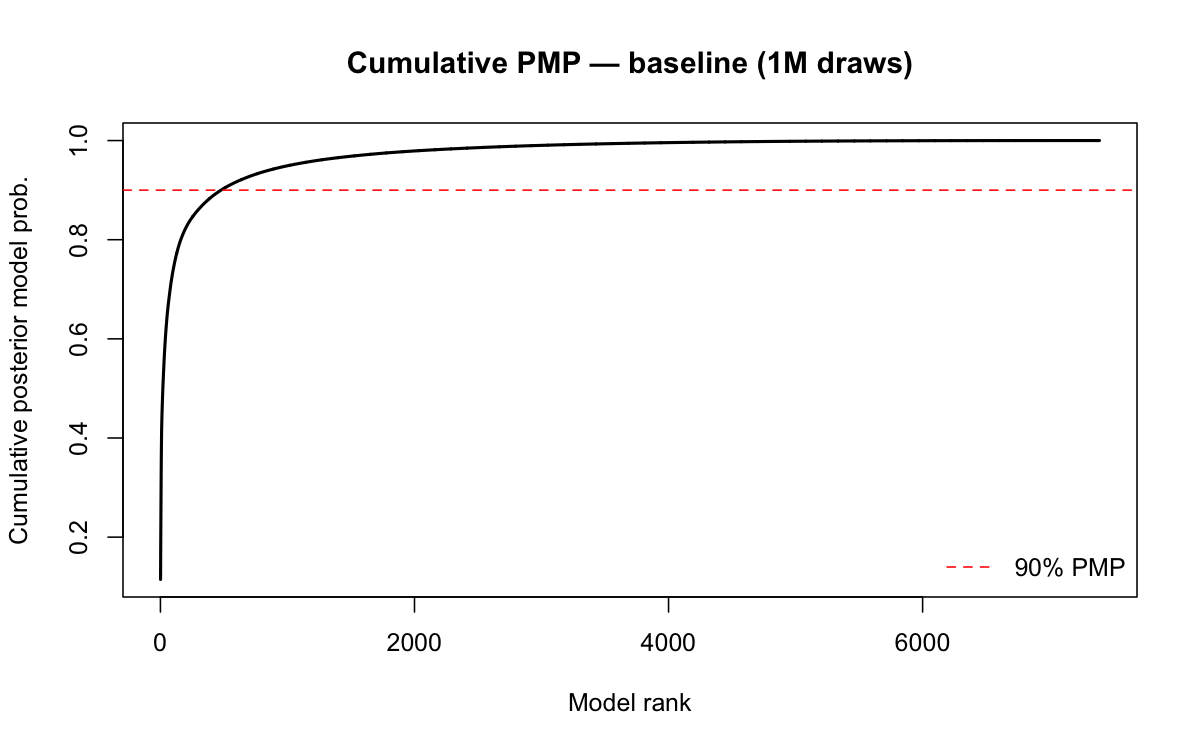}\\[2pt]
        {\footnotesize (b) Cumulative posterior model probability}
    \end{minipage}
    \begin{threeparttable}
        \begin{tablenotes}
            \footnotesize
            \item \textit{Notes:} Panel~(a) depicts the posterior model size distribution and the posterior model probabilities; panel~(b) shows the cumulative posterior model probability across the best models, with the dashed line marking the 90\% threshold. Both refer to the baseline BMA estimation reported in Table \ref{tab:bma1}, using the unit information $g$-prior (UIP) and the dilution model prior due to \citet{eicher2011default} and \citet{george2010dilution}, respectively.
        \end{tablenotes}
    \end{threeparttable}
    \label{fig:posterior1}
\end{figure}

\begin{table}[!]
\centering
\begin{minipage}{0.95\linewidth}
\centering
\caption{Diagnostics of the baseline BMA}
\label{tab:diagnostics1}
\vspace{0.2cm}
\renewcommand{\arraystretch}{1.2}%
\setlength{\tabcolsep}{4pt}%
\begin{tabular}{@{}l@{\hspace{1.5em}}l@{\hspace{3em}}l@{\hspace{1.5em}}l@{}}
\toprule
\multicolumn{2}{@{}l}{\textbf{Sampler \& model space}} & \multicolumn{2}{l@{}}{\textbf{Convergence \& priors}} \\
\cmidrule(lr){1-2}\cmidrule(lr){3-4}
MCMC draws             & $1 \cdot 10^{6}$  & Corr.\ PMP           & 0.9997        \\
Burn-in draws          & $3 \cdot 10^{5}$  & Top models (\% PMP)  & 100\%         \\
Run time               & 6.98 mins         & Shrinkage (avg.)     & 0.9981        \\
Mean no.\ regressors   & 9.4270            & Model prior          & Dilution / 12 \\
Models visited         & 183{,}280         & $g$-prior            & UIP           \\
Model space ($2^{K}$)  & $1.7 \cdot 10^{7}$ & No.\ observations   & 533           \\
\quad of which visited & 1.1\%             &                      &               \\
\bottomrule
\end{tabular}

\footnotesize
{\justifying \textit{Notes:} The table shows technical diagnostic information of baseline BMA estimation reported in Table \ref{tab:bma1}, using the unit information $g$-prior (UIP) and the dilution model prior due to \citet{eicher2011default} and \citet{george2010dilution}, respectively.\par}
\end{minipage}
\end{table}

\begin{table}[!htbp]
\centering
\begin{minipage}{0.95\linewidth}
\centering
\caption{Baseline BMA robustness: alternative priors and Middle East exclusion}
\vspace{0.2cm}
\label{tab:bma2}

\resizebox{\linewidth}{!}{%
\begin{tabular}{lccccccc}
\toprule
 & \multicolumn{3}{c}{\textbf{Prior robustness}} & \multicolumn{3}{c}{\textbf{Subsample robustness}} \\
 & \multicolumn{3}{c}{\textit{(BRIC $g$-prior, random model prior)}} & \multicolumn{3}{c}{\textit{(UIP, dilution; excl.\ Middle East)}}  \\
\cline{2-7}
 & \textit{P. Mean} & \textit{P. SD} & \textit{PIP} & \textit{P. Mean} & \textit{P. SD} & \textit{PIP} \\
\midrule
Standard error (SE) & 2.134 & 0.087 & 1.000 & 2.105 & 0.089 & 1.000 \\
\midrule
\emph{Data characteristics} \\
\hspace*{0.5cm}Sample size                    & 0.035  & 0.011 & 0.960 & 0.031  & 0.013 & 0.897 \\
\hspace*{0.5cm}Data: panel                    & $-$0.165 & 0.063 & 0.936 & $-$0.170 & 0.071 & 0.901 \\
\hspace*{0.5cm}ESG data: LSEG                 & 0.001  & 0.006 & 0.039 & 0.003  & 0.012 & 0.078 \\
\hspace*{0.5cm}Average data year              & 0.001  & 0.005 & 0.058 & 0.000  & 0.001 & 0.006 \\
\hspace*{0.5cm}Average board gender diversity & 0.000  & 0.005 & 0.041 & 0.000  & 0.002 & 0.011 \\
\midrule
\emph{Publication characteristics} \\
\hspace*{0.5cm}Published            & 0.158  & 0.047 & 0.981 & 0.161  & 0.048 & 0.976 \\
\hspace*{0.5cm}Number of citations  & $-$0.032 & 0.007 & 0.997 & $-$0.031 & 0.007 & 0.992 \\
\midrule
\emph{Design of the analysis} \\
\hspace*{0.5cm}Endogeneity control: poor & 0.005  & 0.017 & 0.114 & 0.009  & 0.023 & 0.168 \\
\hspace*{0.5cm}Number of variables       & $-$0.014 & 0.028 & 0.247 & $-$0.003 & 0.013 & 0.055 \\
\midrule
\emph{Control variables} \\
\hspace*{0.5cm}Firm size control          & 0.007  & 0.024 & 0.108 & 0.004  & 0.018 & 0.062 \\
\hspace*{0.5cm}Board independence control & $-$0.001 & 0.006 & 0.035 & $-$0.001 & 0.006 & 0.035 \\
\hspace*{0.5cm}CSR committee control      & $-$0.004 & 0.015 & 0.099 & $-$0.002 & 0.011 & 0.074 \\
\midrule
\emph{Spatial variation} \\
\hspace*{0.5cm}Region: global      & 0.023  & 0.041 & 0.295 & 0.027  & 0.044 & 0.326 \\
\hspace*{0.5cm}Region: Europe      & $-$0.004 & 0.020 & 0.086 & $-$0.002 & 0.012 & 0.053 \\
\hspace*{0.5cm}Region: USA         & $-$0.022 & 0.043 & 0.264 & $-$0.021 & 0.039 & 0.280 \\
\hspace*{0.5cm}Region: Asia        & $-$0.228 & 0.043 & 0.997 & $-$0.225 & 0.042 & 0.994 \\
\hspace*{0.5cm}Region: Middle East & 0.161  & 0.075 & 0.903 & \multicolumn{3}{c}{--- (excluded) ---} \\
\hspace*{0.5cm}Market: emerging    & $-$0.009 & 0.033 & 0.110 & $-$0.007 & 0.028 & 0.073 \\
\hspace*{0.5cm}Sector: financial   & $-$0.001 & 0.008 & 0.033 & $-$0.002 & 0.013 & 0.062 \\
\midrule
\emph{Estimation techniques} \\
\hspace*{0.5cm}Method: linear & 0.022  & 0.033 & 0.361 & 0.026  & 0.036 & 0.404 \\
\hspace*{0.5cm}Method: panel  & $-$0.002 & 0.011 & 0.061 & $-$0.003 & 0.014 & 0.083 \\
\hspace*{0.5cm}Method: IV     & 0.233  & 0.050 & 1.000 & 0.261  & 0.050 & 1.000 \\
\hspace*{0.5cm}Method: GMM    & $-$0.003 & 0.014 & 0.058 & $-$0.002 & 0.013 & 0.059 \\
\midrule
Studies      & 106 & & & 100 & & \\
Observations & 533 & & & 503 & & \\
\bottomrule
\end{tabular}%
}

\footnotesize
{\justifying \textit{Notes:} The table reports two robustness checks on the baseline BMA (Table~\ref{tab:bma1}). The \emph{prior robustness} check (left) re-estimates the full sample with the BRIC $g$-prior \citep{fernandez2001benchmark} and the random model prior \citep{ley2009effect}. The \emph{subsample robustness} check (right) retains the unit information $g$-prior (UIP) and the dilution model prior of \citet{eicher2011default} and \citet{george2010dilution} but excludes Middle East observations, so this regressor drops out of the specification. Both checks reproduce the same set of moderators that cross the PIP $>$ 0.5 threshold in the baseline; the frequentist check is reported once, for the baseline, in Table~\ref{tab:bma1}. Variables are described in Table~\ref{tab:desr_heterogeneity_part1}. P. Mean = Posterior Mean; P. SD = Posterior Standard Deviation; PIP = Posterior Inclusion Probability.\par}
\end{minipage}
\end{table}

\clearpage

\section*{Appendix D}
\thispagestyle{empty}
\vspace{20mm}
\renewcommand{\thetable}{D.\arabic{table}}
\renewcommand{\thefigure}{D.\arabic{figure}}
\setcounter{figure}{0}
\setcounter{table}{0}

\begin{table}[!h]
\centering
\begin{minipage}{\linewidth}
\centering
\caption{Where artificial intelligence was and was not used}
\label{tab:aiuse}
\vspace{0.2cm}
\footnotesize
\begin{tabular}{@{}p{0.24\linewidth}p{0.08\linewidth}p{0.60\linewidth}@{}}
\toprule
Stage & AI used & Detail \\
\midrule
Research question, hypotheses & No & Formulated by the authors. \\
Literature search & No & Google Scholar searches and backward snowballing run by the authors in May and December 2024 (Appendix~A). \\
Screening, inclusion decisions & No & Every inclusion and exclusion decision was made by the authors. \\
Coding of estimates and moderators & No & All 533 estimates were coded by the first author; the share coded by AI is zero. \\
Checking of coded data and reported numbers & Yes & A random sample of coded estimates was re-checked against the primary studies with AI assistance, and the reported numbers were cross-checked against the dataset and re-derived from it as a check. Every discrepancy was resolved by the authors. \\
Estimation, tables, figures & No & All estimates, tables, and figures come from the authors' own Stata and R code in the replication package; the one exception, the PRISMA flow diagram (Appendix~A), was drawn by the authors. \\
Replication package, online appendix & Yes & Assembled with AI assistance and checked by the authors. \\
Language editing & Yes & Editing of text written by the authors. \\
\bottomrule
\end{tabular}

\vspace{0.2cm}
\footnotesize
{\justifying \textit{Notes:} The table gives the disclosure asked for by \citet{Cook2026guidelines} and follows the guiding principles of \citet{Cook2026ai}. The tools used were Claude Opus 4.8, Claude Fable 5, and Claude Sonnet 5 (Anthropic, through Claude Code) and GPT-5.6 Sol (OpenAI, through Codex CLI), between June and July 2026; \texttt{AI\_USE.md} in the replication package records when each tool was used. No search, screening, or coding decision was delegated to AI: the human-coded share is 100\%, so the false-negative rate the guidelines attach to AI-assisted screening does not arise. No figure or table was produced by AI. The coding was done by one author and then re-checked against the primary studies with AI assistance, rather than coded independently a second time, so we report no formal inter-rater statistic; every case the check raised was adjudicated by the authors against the primary study, and the resulting corrections are logged cell by cell in \texttt{Data/CHANGELOG.md}. AI did not influence any analytical choice, such as a specification, weighting scheme, or prior.\par}
\end{minipage}
\end{table}

\end{document}